\documentclass[aoas, preprint]{imsart}

\RequirePackage{amsthm,amsmath,amsfonts,amssymb}
\RequirePackage[authoryear]{natbib}
\RequirePackage[colorlinks,citecolor=blue,urlcolor=blue]{hyperref}
\RequirePackage{graphicx}

\RequirePackage{float, subcaption}
\RequirePackage{multicol, multirow, enumerate}
\RequirePackage{bm, color, comment, tikz}

\startlocaldefs

\newtheorem{lemma}{Lemma}[section]

\newcommand{\bs}{\boldsymbol}
\newcommand{\mbf}{\mathbf}
\newcommand{\mbb}{\mathbb}
\newcommand{\wt}{\widetilde}

\newcommand{\tp}{\intercal}

\newcommand{\JP}[1]{{{\color{red} (revision Jie: #1)}}}
\endlocaldefs

\begin{document}

\begin{frontmatter}
\title{Estimating Fiber Orientation Distribution 
	with Application to Study Brain Lateralization using HCP D-MRI Data}
\runtitle{Estimating FOD with Application to HCP D-MRI Data}

\begin{aug}
\author[A, B]{\fnms{Seungyong} \snm{Hwang}\ead[label=e1]{syhwang@ucdavis.edu}},
\author[A]{\fnms{Thomas} \snm{Lee}\ead[label=e2,mark]{tcmlee@ucdavis.edu}}
\author[A]{\fnms{Debashis} \snm{Paul}\ead[label=e3,mark]{debpaul@ucdavis.edu}}
\and
\author[A]{\fnms{Jie} \snm{Peng}\ead[label=e4,mark]{jiepeng@ucdavis.edu}}
\address[A]{Department of Statistic, University of California, Davis, \printead{e1,e2,e3,e4}}
\address[B]{Department of Genetics, Stanford University}
\end{aug}

\begin{abstract}
	\textit{Diffusion-weighted magnetic resonance imaging} (D-MRI) is an in vivo and non-invasive imaging technology for characterizing tissue microstructure in biological samples. A major application of D-MRI is for white matter fiber tracts reconstruction. 	
	In this paper, we use D-MRI data from the  \textit{Human Connectome Project} (HCP) to study brain lateralization of \textit{Superior Longitudinal Fasciculus} (SLF) -- a major association fiber tract that is involved with motor, visual, spatial, memory and language functions.  Specifically, for each subject, we reconstruct SLF in each brain hemisphere and derive a lateralization score that quantifies hemispheric asymmetry. We then relate this lateralization score to gender and handedness. We find significant handedness effects indicating that SLF lateralization is likely to be different in right-handed and left-handed individuals. 

	Such applications of D-MRI require statistical methods that are computationally scalable to process a large number of images and at the same time  provide accurate estimates of local neuronal fiber structures. 
	In this paper, we propose a computationally efficient method for estimating the \textit{fiber orientation distribution} (FOD) at each brain voxel-based on D-MRI data, referred to as the  \textit{blockwise James-Stein} (\textbf{BJS}) estimator. 
	\textbf{BJS} utilizes the \textit{spherical harmonics} (SH) basis representation of the FOD and adaptively shrinks higher order SH coefficients in a blockwise nonlinear fashion. Through synthetic  experiments, \textbf{BJS} is shown to perform competitively both in terms of computation and accuracy.   
Codes and example scripts for the synthetic experiments and the real data application can be found at \url{https://github.com/vic-dragon/BJS}.
\end{abstract}


\end{frontmatter}

\section{Introduction}
\label{sec:intro}

\textit{Diffusion-weighted magnetic resonance imaging} (D-MRI) is a widely used, non-invasive tool to probe tissue microstructure  of biological samples \textit{in vivo} through measuring  water diffusion characteristics. 
The most important application of D-MRI is the reconstruction of \textit{white matter fiber tracts} -- large axon (a.k.a. nerve fiber) bundles with similar destinations in the brain.  For an example of reconstructed white matter fiber tracts, see Fig. \ref{fig:tract3}.
 D-MRI also has many
 clinical applications, such as detecting brain abnormality in white matter
 due to axonal loss or deformation, which are related to
 many neuro-degenerative diseases including Alzheimer's disease, and surgical
 planning by resolving complex neuronal connections  between white and gray
 matters \citep{nimsky2006implementation}.
 
Mapping white matter fiber tracts is of great importance for studying structural organization of neuronal networks and for understanding brain functionality \citep{Mori07, Sporns11}.  In this paper, we present an application using  D-MRI data from the \textit{Human Connectome Project} (HCP) \citep{VANESSEN2013} to investigate \textit{lateralization (or hemispheric asymmetry)}  of  \textit{Superior Longitudinal Fasciculus (SLF)} -- one of the large lateral association fiber tracts located in each hemisphere involved in motor, visual, spatial, memory and language functions \citep{Makris2004}. 

Brain  lateralization refers to the tendency for some neural functions to be specialized to one hemisphere of the brain. 
The most well-known example is the lateralization of the language pathway and it has been studied recently through neuroimaging technologies including D-MRI and functional MRI (fMRI). For example,  \cite{Catani2007} and \cite{Gharabaghi2009} investigated the perisylvian language pathway, the direct connections between Broca’s and Wernicke’s territories, through D-MRI tractography. \cite{houston2019} investigated the association of diffusion tensor imaging (DTI) metrics  with language function and demographic features including age and gender by making use of \textit{tract based spatial statistics (TBSS)} \citep{SMITH2006}.  \cite{SZAFLARSKI2012} investigated language lateralization in left-handed children through fMRI. Lateralization of other pathways, such as the motor pathway, has also been studied through imaging technologies. E.g., \cite{Seizeur2014} studied the association between corticospinal tract asymmetry and handedness through D-MRI tractography.

In the following, we first briefly describe the D-MRI technology and the HCP D-MRI data. We then discuss various D-MRI models  and estimators. Finally, we briefly describe the application and findings,  as well as  the data analysis pipeline. 

In biological samples, water diffuses preferentially along tissue structures which leads to  \textit{anisotropic diffusion}. For example,  in the brain water tends to diffuse faster along  white matter fiber tracts.  In a D-MRI   experiment, multiple magnetic field gradients, each along a  \textit{gradient direction} (represented by a unit-length vector in $\mathbb{R}^3$),
 are successively applied to the tissue sample. For each gradient application, the corresponding signal at a given voxel,  is sensitized with the amount of water diffusion within this voxel. Specifically,  the greater amount of diffusion along the gradient direction, the lower is the signal intensity (or the greater signal attenuation) due to the greater degree of phase disruption.  This is the reason that D-MRI measurement is often referred to as  \textit{diffusion weighted measurement}. 
 
 Suppose at a particular voxel, water mainly diffuses along the left-right direction. Then  for  gradient applications perpendicular to the left-right direction (e.g., those along the superior-inferior or anterior-posterior directions), there would be little signal attenuation as there is little  water motion along their directions. 
 Consequently, at this voxel, the signal intensity corresponding to such gradient applications would be (nearly) the same as the baseline signal intensity (induced by a strong constant background magnetic field, referred to as the $b_0$- field). On the other hand, if a gradient application is along the left-right direction, then there would be high signal attenuation such that at this voxel, the signal intensity corresponding to this gradient application would be (much) smaller than the baseline signal intensity.  

In addition to the direction of gradient application, the amount of signal attenuation  is also affected by other factors including  field
strength and the duration of gradient application. The aggregated effect of these factors is  reflected by an experimental parameter  called the \textit{$b$-value}.   In short,  the higher the $b$-value, the greater is the amount of signal attenuation and the more sensitive are the measurements  to  water diffusion.
For more details of the D-MRI technology and data acquisition, the readers are referred to  \cite{LeBihanEtal2001,Mori07,Jones2010}.


D-MRI data from HCP have diffusion weighted measurements taken under  three different \textit{b}-values ($1,000s/mm^2, 2,000s/mm^2, 3,000s/mm^2$). For each \textit{b}-value,  at each voxel (size: $1.25\times 1.25\times 1.25 mm^3$) 
on a $145\times 174\times 145$ 3D brain grid,   there are measurements corresponding to a common set  of $90$ distinct gradient directions.   
 Moreover, 6 non-diffusion weighted images  (referred to as  \textit{$b_0$ images}) are obtained under the constant background magnetic field. In summary, for each of the three \textit{b}-values, a  HCP D-MRI data set consists of $96$ grey scale images on a  $145\times 174 \times 145$ 3D grid, along with $90$ 3D unit vectors representing the $90$ gradient directions. 


One of the earliest and still widely used D-MRI model  is the  \textit{single tensor model} where at each voxel, the diffusion process is modeled by a 3D Gaussian distribution whose covariance matrix is referred to as the \textit{diffusion tensor} \citep{Mori07}. The single tensor model is the reason that D-MRI is often referred to as \textit{diffusion tensor imaging (DTI)}. 
 
 Specifically, 
the (noiseless) diffusion weighted signal at voxel $\mbf{v}$  along a gradient direction $\mathbf{x}$ is given by : 
\begin{equation}
\label{eq:single-tensor}
S(\mbf{v}, \mbf{x}) = S_0(\mbf{v}) \exp\left\{-b\mbf{x}^\tp {\mbf{D}}(\mbf{v}) \mbf{x}\right\},
\end{equation}
where $\mbf{x}$ is a 3D unit-length vector, $S_0(\mbf{v})$ is the non-diffusion-weighted signal intensity at voxel $\mbf{v}$, $\mbf{D}(\mbf{v})$ is a
$3\times 3$ positive definite matrix -- the diffusion tensor at voxel $\mbf{v}$, and  $b>0$ is the \textit{$b$-value}.    
As a tensor has only six parameters, the single tensor model can be fitted with as few as seven diffusion measurements.

Note that, if the gradient direction $\mbf{x}$ is aligned with the principal eigenvector of the tensor $\mbf{D}(\mbf{v})$, then we will observe the most signal attenuation. Thus, the voxel-level fiber/diffusion orientation is extracted as  the principal eigenvector of the (estimated) diffusion tensor and used as inputs in \textit{tractography} algorithms for white matter tracts reconstruction  \citep{Basser-Pajevic-Pierpaoli00}.

The single tensor model also provides some useful image contrasts, most notably, the \textit{fractional anisotropy (FA)} that quantifies the degree of anisotropic diffusion at a voxel: 
\begin{equation}
	\label{eq:FA}
	FA:=\sqrt{\frac{1}{2}} \frac{\sqrt{(\lambda_1-\lambda_2)^2+(\lambda_2-\lambda_3)^2+(\lambda_3-\lambda_1)^2}}{\sqrt{\lambda_1^2+\lambda_2^2+\lambda_3^2}},
	\end{equation}
where $\lambda_1 \geq \lambda_2 \geq \lambda_3>0$ are the three eigenvalues of the diffusion tensor $\mbf{D}$ at that voxel. When diffusion is isotropic, i.e.,  $\lambda_1=\lambda_2=\lambda_3$, the FA value would reach the lower limit $0$; Whereas when diffusion is highly anisotropic, i.e, $\lambda_1 >>\lambda_2,\lambda_3$, the FA value would approach the upper limit $1$.

Despite its simplicity and usefulness, the single tensor model cannot resolve \textit{ intravoxel orientational heterogeneity } (a.k.a. crossing fibers) -- multiple fiber populations with distinct orientations within a voxel. This is estimated to be present in approximately 30\% white-matter voxels, and so,  in such regions,  any single tensor model would lead to misleading  FA values  and poor direction estimation  which adversely affects fiber reconstruction.

\begin{figure}[H]
	\centering
	\includegraphics[width=\textwidth]{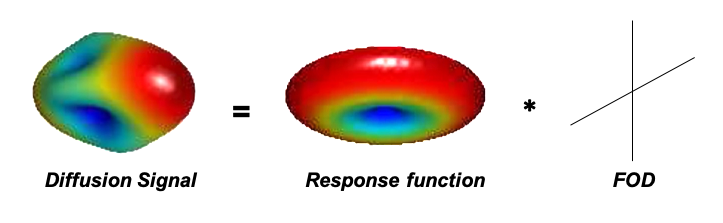}
	\caption{\textbf{FOD model of D-MRI signal:} at a voxel with two fibers crossing  at a  $60^{\circ}$ \textit{separation angle} .}
	\label{fig:fodmodel}
\end{figure}

The SLF in our application  is known to be difficult to reconstruct due to crossing between fibers of SLF and the corticospinal tract (CST) \citep{CATANI2010}. 
This motivates us to employ an alternative  model that expresses the D-MRI signal at each voxel as  a convolution of an underlying \textit{fiber orientation distribution (FOD)} function and an axially symmetric response function \citep{tournier2004}.  See Fig. \ref{fig:fodmodel} for a graphical illustration at a voxel with two fiber bundles crossing at $60^{\circ}$. 

The FOD model directly describes the local spatial arrangement of axonal fiber bundles
and thus is particularly attractive when  the downstream goal is white matter fiber tracts reconstruction. 
Particularly, the FOD model is able to resolve crossing fibers within a voxel at the expense of requiring \textit{high angular resolution diffusion imaging (HARDI)} data \citep{TuchEtal02, Hosey-Williams-Ansorge05}, where a large number of gradients (such as in the HCP D-MRI experiments) is sampled.

Since  FOD describes the distribution of fiber bundle orientation at each voxel, it is reasonable to think of the FOD as a smooth 
function with a few sharp peaks, where each peak corresponds to a distinct major fiber bundle within the voxel; or no peak at all in case of isotropic diffusion.   Once the FOD is estimated, the peak directions can be extracted and then used as inputs for tractography algorithms. This means that, in presence of any fiber bundle in the voxel,  it is imperative for the estimators to have sufficient \textit{angular resolution}, i.e., sharp peak(s).

In this paper, we propose a \textit{blockwise James-Stein} type estimator, referred to as \textbf{BJS}, for FOD estimation at each voxel.  Through extensive synthetic experiments, we compare \textbf{BJS} with two other FOD estimators, namely, \textbf{SHridge} which uses a ridge-type penalty \citep{maxime2006, YanCPP2018}, and \textbf{SCSD}  which applies an iterative super-resolution sharpening \citep{tournier2007} upon the \textbf{SHridge} estimator. The results demonstrate that \textbf{BJS} achieves competitive performance in terms of direction estimation, particularly in maintaining  angular resolution.  It  is also at least 10 times faster than the other two methods. 
The  computational efficiency of \textbf{BJS} is important for the kind of  applications as  in this paper, where
the FOD  model needs to be  fitted on  a large number of voxels per image  (here $\sim 100K$)  for a large number of images (here $\sim 200$).

In this paper, we use the D-MRI data from $184$ (unrelated) HCP young adult subjects to investigate the association between left- and right-   SLF  asymmetry (measured by a lateralization score) with gender and handedness -- two commonly considered demographic/behavioral features in brain lateralization studies. 
We find significant handedness main effects on the lateralization score indicating that SLF lateralization pattern is likely to be different between right-handed and left-handed individuals. On the other hand, there is no significant gender main effect or gender-handedness interaction effect.

\begin{figure}[H]
	\centering
	\includegraphics[width=\textwidth]{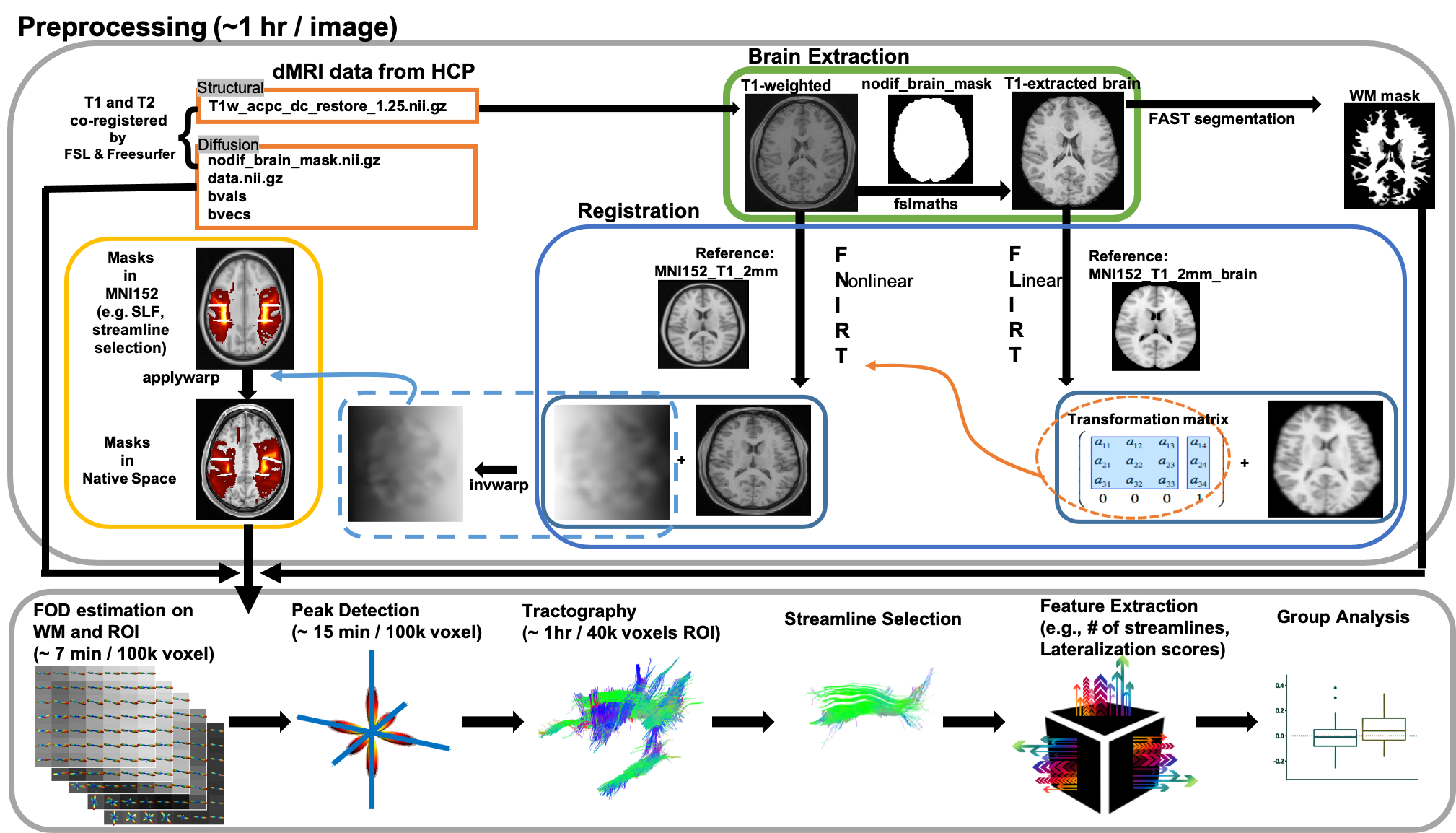}
	\caption{\textbf{HCP D-MRI application: Schematic plot of the data analysis pipeline.}}
	\label{fig:pipeline}
	{\tiny (processing time is under a Xeon 72 core, 2.3GHz, 256GB RAM linux server)}
\end{figure}

To carry out this application, we develop a data analysis pipeline that can be adopted to investigate associations between other D-MRI derived brain structural connectivity features and external variables. The pipeline is illustrated in Fig. \ref{fig:pipeline} and includes the following major steps:  \textbf{Preprocessing} -- conducting brain extraction, white matter segmentation and registration  using the software \textit{FSL} version 6.0.0 \citep{JENKINSON2012} and R packages  \textit{fslr} \citep{fslr} and \textit{neurohcp} \citep{neurohcp} from the \textit{neuroconductor} repository; \textbf{SLF masks} --  creating masks using \textit{FSLeyes} \citep{fsleyes} and the \textit{JHU White-Matter Tractography Atlas} \citep{WAKANA2007,HUA2008}; \textbf{FOD estimation and peak detection} -- deriving \textbf{BJS} estimates for the white-matter voxels within the  SLF masks; Extracting the peaks of the estimated FODs by a peak detection algorithm \citep{YanCPP2018};
\textbf{SLF reconstruction} -- using extracted peak directions as inputs in a \textit{deterministic tractography algorithm} – \textit{DiST} \citep{WongLPP2016}; Applying streamline selection to further improve the reconstruction; \textbf{Feature extraction} --
computing a lateralization score defined as the relative difference between the numbers of fibers (streamlines) in the left- and right-hemisphere reconstructed SLF; \textbf{Group analysis} -- relating the lateralization score with gender and handedness through a two-way ANOVA model.

The main contributions of this paper are threefold. First, we propose a novel computationally scalable FOD estimator, \textbf{BJS}, which enriches the set of statistical tools for analyzing D-MRI data. Second, we establish a D-MRI data analysis pipeline that can be used for studying associations between D-MRI derived brain structural connectivity features and external variables (such as demographic and behavioral features or cognitive measures). Third, we find significant handedness effects in SLF lateralization which enhances our understanding of brain lateralization and brain function. 

The rest of the paper is organized as follows.  Section~\ref{sec:method} describes the FOD model and the proposed \textbf{BJS} estimator. In Section~\ref{sec:simul}, \textbf{BJS} is compared with two competing methods through synthetic experiments.  Section~\ref{sec:application} reports the HCP D-MRI application on SLF lateralization.  We conclude the paper with a discussion in Section~\ref{sec:discussion}. Further details can be found in the Supplementary Material.

\section{Methodology}
\label{sec:method}

In this section, we first introduce the \textit{fiber orientation distribution (FOD)} model  \citep{tournier2004}, followed by a discussion of  two existing FOD estimators, referred to as  \textbf{SH-ridge} \citep{maxime2006, YanCPP2018} and \textbf{SCSD} \citep{tournier2007}. We then describe the proposed estimator \textbf{BJS}, which achieves superior or similar performance as  \textbf{SH-ridge} and \textbf{SCSD}, albeit being computationally much more efficient.

\subsection{FOD Model and SH Representation}
 The FOD model assumes that the diffusion signal at each voxel is a \textit{spherical convolution} between the \textit{response function} --  an axially symmetric \textit{convolution kernel} that characterizes water diffusion when there is a single dominant fiber bundle aligned with the z-axis (the bottom-top axis), and the FOD -- a probability density function defined on the unit sphere -- of fiber bundle orientation at that voxel.

Specifically, in the FOD model,  the (noiseless) diffusion weighted signal at voxel $\mbf{v}$  along a gradient direction $\mathbf{x}$ is given by:
\begin{equation}\label{eq:dMRI_signal} 
S(\mbf{v}, \mbf{x}) = \int_{\mbb{S}^2} R(\mbf{x}^T \mbf{y}) F(\mbf{v}, \mbf{y}) d\omega(\mbf{y}), \qquad \mbf{x} \in \mbb{S}^2,
\end{equation}
where $d\omega(\mbf{y})$ is the volume element of the 3D unit sphere $\mbb{S}^2$; 
$F(\mbf{v}, \cdot)$ is a symmetric spherical probability density function; and $R(\cdot)$ is an axially symmetric  kernel and is assumed to be the same across voxels and fiber bundles.
In practice, we can estimate  $R(\cdot)$ based on voxels with a single dominant fiber bundle, characterized by high FA values under the single tensor model. It is shown in \cite{YanCPP2018} that the FOD model is quite robust to the specification of the response function. 
See Section \ref{sec:hcp-fod} and the Supplementary Text (Section \ref{sec:supp-response}) for details.
In the following, $R(\cdot)$ is assumed to be known.  Moreover, to simplify notation, hereafter, we suppress the dependency on voxel index $\mbf{v}$ in $S$ and $F$. 
Our goal is to estimate the FOD  $F(\cdot)$ based on the observed diffusion measurements. As can be seen from model (\ref{eq:dMRI_signal}), this amounts to a \textit{spherical deconvolution} problem.

Since $S(\cdot), F(\cdot)$ and $R(\cdot)$ are real and symmetric functions, they can be represented by \textit{real symmetrized spherical harmonic (SH) basis}.  
Let $\{\wt\Phi_{lm}: -l \leq m \leq l; l=0,1,\ldots\}$ denote the complex SH basis. 
Then, a real symmetrized SH basis is defined as \citep{DescoteauxAFD07}: 

\begin{equation}\label{eq:symm_real_SH}
{\Phi}_{lm} = \begin{cases}
\sqrt{2} ~\cdot~ \Re(\wt\Phi_{lm}), & \mbox{if}~ - l \leq m < 0 \\
\Phi_{l0}, & \mbox{if}~ m = 0 \\
\sqrt{2} ~\cdot~ \Im(\wt\Phi_{lm}), & \mbox{if}~ 0 < m \leq l \\
\end{cases}
\end{equation}
for $l=0,2,4,\ldots$, where $\Re(z)$ and $\Im(z)$ denote the real and imaginary parts of $z \in \mbb{C}$, respectively. 

Since the response function $R(\cdot)$ is axially symmetric, $\langle R, \Phi_{lm}\rangle = 0$ unless $m = 0$. Let $r_l = \langle R, \Phi_{l0}\rangle$ and $f_{lm} = \langle F, \Phi_{lm}\rangle$ be the SH coefficients of $R(\cdot)$ and $F(\cdot)$, respectively. 
Then, by equation (\ref{eq:dMRI_signal}), the D-MRI signal $S(\cdot)$ has SH coefficients: 
\begin{equation}\label{eq:dMRI_signal_SH_coeff}
s_{lm} = \langle S, \Phi_{lm}\rangle = \sqrt{\frac{4\pi}{2l+1}} r_l f_{lm}, -l\leq m \leq l; ~~l=0,2,\ldots
\end{equation}

The observed D-MRI measurements $\{y_i\}_{i=1}^n$  are noise corrupted versions of $S(\cdot)$ measured along $n$  gradient directions $\{\mbf{x}_i\}_{i=1}^n$.
The major source of noise (after removing artifacts due to eddy currents, echo planar imaging distortion and subject motion) in D-MRI data is the thermal noise in the MRI scanner. It is modeled as independent and additive white noise on the real and imaginary parts of the signal. Since the observed D-MRI measurements are $\ell_2$ norms of the complex-valued signal from the MRI scanner, they follow a \textit{Rician distribution} \citep{GudbjartssonP95}. However, when the \textit{signal-to-noise ratio(SNR)} level is high as is the case of HCP D-MRI data (see Fig. \ref{fig:snr}), Rician noise can be approximated by additive Gaussian noise \citep{CarmichaelCPP2013}.

Under the assumption that a finite level (up to $l_{\max}$) of SH basis can represent $S(\cdot)$ and $F(\cdot)$, the D-MRI measurements can then be modeled as:
\begin{equation}\label{eq:dMRI_data_SH}
  \mbf{y} = \bs\Phi \mbf{R} \mbf{f} + \bs\varepsilon\textnormal{,}~~~~ \bs\varepsilon \sim N(0,\sigma_{\varepsilon}^2\mbf{I}_n),
\end{equation}
where $\mbf{R}$ is an $L\times L$ diagonal matrix with $l$-th diagonal block equal to $r_l \sqrt{4\pi/(2l+1)}\mbf{I}_{2l+1}$, for $l=0,2,\ldots,l_{\max}$ and $L=\frac{(l_{\max}+1)(l_{\max}+2)}{2}$ is the total number of SH basis; and $\bs\Phi$ is the $n\times L$ matrix with the elements in the $i$-th row and $(l,m)$-th column given by ${\Phi}_{lm}(\mbf{x}_i)$; and $\mbf{f}=(f_{lm})$ is the $L\times 1$ vector of SH coefficients of the FOD $F(\cdot)$. Moreover,  the vector $\bs\varepsilon = (\varepsilon_i)_{i=1}^n$ represents observational noise and is assumed to have independent coordinates that follow a Gaussian distribution with $\mbb{E}(\varepsilon_i)=0$ and Var$(\varepsilon_i) = \sigma_\varepsilon^2$. 

 In order to achieve sufficient angular resolution, higher order spherical harmonics are needed to represent the FOD (i.e.,   sufficiently large $l_{\max}$). In practice, we choose $l_{\max}$ to be the largest even number such that the sample size $n$ is still greater than the number of SH basis $L$.  Specifically, in the synthetic experiments $l_{\max} = 6, 10, 12$ ($L=28,66, 91$) for $n=41, 91, 321$ gradient directions, respectively, and  in the HCP application, $l_{\max}=10$ ($L=66$) is used as there are $n=90$ gradient directions.

 Also note that, due to the  decay of the singular values of the design matrix resulting from the decrease of higher order SH coefficients of the response function $R(\cdot)$,  deconvolution becomes increasingly unstable and more susceptible to noise amplification when higher order harmonics are used in FOD representation.   
Therefore, appropriate regularization is the key to accurate FOD estimation and fiber direction extraction. 

In the following, we first describe two existing FOD estimators:  \textbf{SH-ridge}  which is based on a ridge-type linear regularization; and \textbf{SCSD}  which is based on an iterative \textit{sharpening process}  for improving \textbf{SH-ridge} through highlighting the (estimated) peaks. 
We then propose a \textit{blockwise James-Stein type shrinkage estimator}, referred to as \textbf{BJS}. Blockwise nonlinear shrinkage strategies have been used in the literature for adaptive estimation in nonparametric regression \citep{Cai1999,CaiLZ2009}  and linear inverse problems \citep{CavalierT2001,CavalierT2002}, under the setting of i.i.d. noise. 
Since FOD is expected to be mostly a smooth function with a few sharp peaks (or no peak at all), we expect relatively smaller coefficients for higher level harmonics. Therefore, a nonlinear shrinkage procedure with a blockwise adaptive choice of the shrinkage parameter  is expected to be more efficient than linear estimators such as \textbf{SH-ridge}.

\subsection{ SHridge and SCSD}



The \textbf{SHridge} estimator is motivated by  \cite{maxime2006} who proposed the \textit{Laplace-Beltrami} regularization to estimate the \textit{orientation distribution function (ODF)} \citep{Tuch02, Tuch04}. The same penalty can be used for FOD estimation \citep{YanCPP2018}:
\begin{equation}\label{eq:SHridge}
	\min_f ||\mbf{y} - \bs\Phi\mbf{R}\mbf{f}||^2_2 + \lambda\mathbb{E}(F), ~~~ \mathbb{E}(F):= \int_{\Omega} (\bigtriangleup_b F)^2 d\Omega = \mbf{f}^T\mbf{P}\mbf{f},
\end{equation}
where $\mbf{P}$ is a block diagonal matrix with entries $l^2(l+1)^2$ and block size $2l+1$ for $l=0,2,\dots,l_{\max}$; $\bigtriangleup_b$ is the spherical Laplacian operator, and $\mathbb{E}(F)$ is a measure of roughness of spherical functions.
With the objective function (\ref{eq:SHridge}), the estimated coefficients of FOD are: 
\begin{equation}\label{eq:SHridge_est}
	\hat{\mbf{f}}^{SHridge} = (\mbf{R}\bs\Phi^T\bs\Phi\mbf{R}+\lambda\mbf{P})^{-1}\mbf{R}\bs\Phi^T\mbf{y}. 
\end{equation}
The tuning parameter $\lambda$ can be chosen by a grid search and the Bayesian information criterion (BIC) \citep{Schwarz1978}.

The \textbf{SHridge} estimator suffers from low angular resolution and is inaccurate when there are crossing fibers with moderate to small crossing angles (see Section \ref{sec:simul}).  One strategy to improve  the angular resolution  of FOD estimator is through a \textit{sharpening process} which makes the major peak(s) more prominent and at the same time suppresses small peaks as these are more likely due to noise.

 Specifically, \cite{tournier2007} proposed a \textit{sharpening} procedure,  referred to as the \textit{superCSD}, 
which iteratively suppresses small (including negative) values and elevates
large values through a super-resolution SH representation (with an order $l_{\max}^s \geq l_{\max}$). The detailed procedure can be found in the Supplementary Text (Section \ref{sec:supp-scsd}).

We refer to the estimator resulting from applying the \textit{superCSD} procedure to the \textbf{SHridge} estimator as \textbf{SCSD}.  Although \textbf{SCSD} is able to improve upon \textbf{SHridge} (see Section \ref{sec:simul}), it dose so at the expense of considerable extra computational overhead. Next, we propose a new estimator \textbf{BJS} that is able to achieve a similar angular resolution as \textbf{SCSD}, albeit with much less computational  cost.

\subsection{Blockwise James-Stein Shrinkage Estimator (BJS)}

When the sample size $n$  is greater than the number of SH basis $L$, then (i) $\bs\Phi^T\bs\Phi$ is well-conditioned 
(assuming a weak requirement on the configuration of the gradients directions, e.g., from an \textit{icosphere mesh}); and (ii) $\mbf{R} \bs\Phi^T \bs\Phi \mbf{R}$ is nonsingular. 
However, due to finite sampling,  $\bs\Phi^T\bs\Phi$ is not an identity matrix. Also, for larger $L$, the matrix $\mbf{R}$ and consequently, the matrix $\mbf{R}\bs\Phi^T\bs\Phi \mbf{R}$ becomes significantly ill-conditioned, since $r_l$ decreases to zero as $l$ increases. Therefore, a linear estimator of FOD (e.g. the \textbf{SHridge} estimator) is likely to be inefficient. This motivates us to consider a nonlinear shrinkage procedure, referred to as \textbf{BJS}.

Specifically, we first obtain the ordinary least squares (OLS) solution, referred to as the  \textit{transformed observations}. 
We then  partition the transformed observations into blocks corresponding to the frequency levels of the SH basis and apply a James-Stein type shrinkage estimator  within each block.  
Since the SH transform of the response function is constant within each harmonic frequency level, the covariance matrix of the transformed data is reasonably homogeneous and well-conditioned within each  block. Moreover, inspired by \cite{laurent2000} and \cite{CavalierT2001}, we adopt a more heavily penalized version of the James-Stein shrinkage which accounts for non-isotropic covariance of the observations, thus allowing for \textit{heteroscedasticity} as well as dependency among the observations. Finally, we employ a post-estimation \textit{one-step} super resolution sharpening to enhance the localized peaks of the estimated FOD.  Note that,  \textbf{BJS}  does not involve any grid search or iteration, thus it is computationally much more efficient than \textbf{SHridge} and \textbf{SCSD}, and scales well for processing a large number of diffusion images.

The detailed \textbf{BJS} procedure is as follows: 

\subsubsection*{Step 1: Transformation}

Multiply $\mbf{K} = {\mbf{R}}^{-1}(\bs \Phi^T \bs\Phi)^{-1} \bs\Phi^T$ to both sides of (\ref{eq:dMRI_data_SH}) to obtain the transformed 
observations: 
\begin{equation}\label{eq:Preconditioning}
\mbf{z} = \mbf{K}\mbf{y} = \mbf{f} + \mbf{K}\varepsilon,
\end{equation}

where
\[
  \mbox{Var}(\mbf{z}) = \sigma^2_{\varepsilon}\mbf{K}\mbf{K}^T = \sigma^2_{\varepsilon}{\mbf{R}}^{-1}(\bs\Phi^T\bs\Phi)^{-1}{\mbf{R}}^{-1}: = \sigma^2_{\varepsilon}\mbf{V}.
\]

\subsubsection*{Step 2: Blockwise James-Stein Shrinkage}

In this step, we estimate $ \mbf{f} $ through a blockwise James-Stein type estimator by applying an adaptive nonlinear shrinkage within each block. 
 
Denote the block of $ \mbf{f} $ and $\mbf{z}$ corresponding to the $l$-th level SH basis by $ \mbf{f} ^{(l)}$ and $\mbf{z}^{(l)}$, respectively. The $l$-th block consists of $(2l+1)$ coordinates of the respective vector for $l = 0, 2, 4, ..., l_{\max}$ and there are $1+ \frac{l_{\max}}{2}$ total blocks. Let $\mbf{V}^{(l)}$ be the corresponding $(2l+1)\times(2l+1)$ submatrix of $\mbf{V}$ and $\bs\eta^{(l)}$ be the $l$-th block of the transformed noise vector $\bs \eta :=\mbf{K}\varepsilon$. 

Note that, 
$\bs\eta^{(l)}$ follows $N(0,\sigma^2_{\varepsilon}\mbf{V}^{(l)})$. Moreover, $\mbf{V}^{(l)}$ equals $\frac{2l+1}{4 \pi r_l^2}$ multiplying  the corresponding $(2l+1)\times(2l+1)$ submatrix of  $(\bs\Phi^T\bs\Phi)^{-1}$. This is because within each  block $\mbf{R}$ is  a scalar multiple of the identity matrix.  Consequently, $\mbf{V}^{(l)}$  is much better conditioned than $\mbf{V}$ and this motivates the  blockwise shrinkage estimator described below.

For each level $l$, we have $\mbf{z}^{(l)} =  \mbf{f} ^{(l)} + \bs\eta^{(l)}$. 
For $l=0, 2, \ldots ,l_0$ with $l_0 \geq 2$ a prespecified even number,  $\hat{\mbf{f} }^{(l)} := \mbf{z}^{(l)}$. For $l >l_0$,  we adopt a modified version of James-Stein shrinkage described in (\ref{eq:BlockJS_est}),  which accounts for the non-isotropic covariance of $\bs\eta^{(l)}$:  

\begin{align}\label{eq:BlockJS_est}
  \hat{ \mbf{f} }^{(l)} = \left( 1-\frac{\hat{\sigma}^2_{\varepsilon}(\|\bs\lambda_l\|_1 + 2\|\bs\lambda_l\|_2 \sqrt{t^{(l)}} + 2\|\bs\lambda_l\|_{\infty}t^{(l)} )}{\|\mbf{z}^{(l)}\|^2_2}\right)_{+}\mbf{z}^{(l)},  ~~~ l > l_0, 
\end{align}
where $\bs\lambda_l$ is the vector of ordered eigenvalues of $\mbf{V}^{(l)}$,
$
\|\bs\lambda_l\|_1, ~ \|\bs\lambda_l\|_2, ~ \|\bs\lambda_l\|_{\infty}$ are $\ell_1 , \ell_2, \ell_\infty$ norm of   $\bs\lambda_l$, respectively, 
 $t^{(l)} = 2\log(2l + 1)$ is a regularization parameter. The error variance $\sigma^2_{\varepsilon}$ is estimated by the mean squared error (MSE) of OLS: 
\[
\hat{\sigma}^2_{\varepsilon} = \frac{\|\mbf{y} - \bs\Phi(\bs\Phi^T\bs\Phi)^{-1}\bs\Phi\mbf{y}\|}{n-\mbox{rank}(\bs\Phi)}.
\]
Note that, even with an ill-conditioned system, we can still get a good fit of the observations by OLS and consequently a good estimate of the error variance.

Note that, shrinkage is only applied to SH coefficients with a level higher than $l_0$, whereas low order SH coefficients are estimated by OLS. The reason is to avoid excessive bias as low order SH coefficients are expected to be large. Specifically, $l_0$ should be an even number no less than $2$, which limits the possible choices for $l_0$. In this paper, we set $l_0=4$, meaning that we do not shrink the first three levels (i.e., $l=0,2,4$) of SH coefficient estimates.

	The adaptive nonlinear shrinkage factor in (\ref{eq:BlockJS_est}) is determined by making use of a  tail-probability bound for quadratic forms of Gaussian vectors from \cite{laurent2000}
	(see Lemma \ref{lemma:quad_from_tail_bound} in the Supplementary of Text)
	as well as the representation, 
	\[
	\|\bs\eta^{(l)}\|_2^2 \sim \sigma_{\varepsilon}^2\sum_{i=1}^{2l+1} \lambda_{l,i} w_{l,i}^2,
	\]
	where $\lambda_{l,i}$ is the $i$-th largest eigenvalue of $\mbf{V}^{(l)}$, and $w_{l,i} \sim N(0,1)$.

	Moreover, the shrinkage parameter $t^{(l)}$  is  of  the form  $c\log(2l+1)$  where $c>1$ is a constant factor. This form ensures that, 
	within a block, the probability of falsely detecting a (nonexistent) signal goes to zero at the rate $(2l+1)^{-c}$  (by Lemma  \ref{lemma:quad_from_tail_bound}). 
	Moreover, if there is no signal at all, then with probability tending to one (as $l  \rightarrow \infty$), except for within at most a finite number of blocks, the parameters will be shrunk to zero  (by the \textit{Borel-Cantelli lemma}). Particularly,  for blocks corresponding
		to higher SH levels, larger shrinkage is applied so that the noise is more aggressively suppressed.
		In this paper, we set $c=2$. Based on  the results of a sensitivity experiment (Table \ref{table:sensitivity_degree45}), \textbf{BJS} estimator is quite robust with respect to $c$.  On the other hand,  too small $c$ values could fail to suppress enough noise, whereas very large  $c$ values could suppress too much signal and lead to loss of sharp
		features. The suggested value $c=2$  achieves a good
		balance between noise suppression and feature retention.

\subsubsection*{Step  3: Post-estimation Sharpening}

The evaluation of the estimated FOD on an arbitrary grid is given by 
$
\hat{F}= \tilde{\mbf{\Phi}} \hat{f},
$
where $\tilde{\mbf{\Phi}}$ is an $\tilde{n} \times L$ matrix representing the evaluation of the $L$ SH basis on this grid. 
Since the estimated FOD $\hat{F}$ may have negative values caused by artificial oscillations, to impose nonnegativity, the SH coefficients $\hat{f}$ is further updated through a one-step super-resolution sharpening process using an $l^s_{\max} (\geq l_{\max})$ order SH representation, which not  only suppresses negative values but also sharpens the peak(s). In this paper, $l_{\max}^s=16$ is used for the synthetic experiments with separation angle $30^{\circ}$; for all other cases and the HCP application, $l_{\max}^s=12$ is used. See Table \ref{table:sensitivity_lmaxs} for a sensitivity experiment on the effect  of $l_{\max}^s$.  
Details of the one-step sharpening are available in the Supplementary Text (Section \ref{sec:supp-bjs}).

Note that, both \textbf{BJS} and \textbf{SCSD} use a nonlinear super-resolution sharpening process. However, there are two differences. First,   \textbf{BJS} uses super-resolution sharpening only once, whereas  \textbf{SCSD}  uses a computationally much more costly iterative procedure. Second, 
	 in \textbf{BJS},  only negative values are suppressed, whereas in \textbf{SCSD} both negative and small positive values are suppressed.  
	Suppressing small positive values helps the iterative  procedure to converge, but this is not necessary for the one-step procedure. In a sensitivity experiment, we examine the effect of suppressing small positive values in the one-step sharpening of \textbf{BJS} and find that it does not lead to better estimation (see Table \ref{table:sharpening}). 
	
	A good initial estimator is crucial for the success of the one-step sharpening process. This is evidenced by the comparison between \textbf{BJS} and \textbf{SCSD} (Section \ref{sec:simul}), where \textbf{BJS} is able to achieve comparable or better accuracy in FOD estimation largely due to its superior initial estimator to  \textbf{SHridge} (which is used as the initial estimator in \textbf{SCSD}).

\subsubsection*{Further discussion on the James-Stein shrinkage factor}

Our formulation of the blockwise shrinkage factor in (\ref{eq:BlockJS_est})  explicitly accounts for heteroscedastic and correlated noise within each block. In particular,
one important difference of the current setting from those in the existing works is that,  the covariance matrix of the transformed data is non-diagonal. This is because, due to finite sampling, the design matrix does not diagonalize in the same SH basis as the convolution kernel. Addressing this point requires a careful calibration of the shrinkage factor in  \textbf{BJS}.

\cite{CavalierT2002} solved a linear inverse problem under a Gaussian sequence model with i.i.d. noise, using a blockwise James-Stein shrinkage rule. They first converted the inverse problem to a direct estimation problem with independent but heteroscedastic noise. 
They showed that a larger value of the blockwise shrinkage factor than in the ordinary James-Stein shrinkage procedure gave better control on the variance of the estimator at the expense of slightly increased bias. In our context, we also deal with an inverse problem, 
so a good balance between variance and bias of the estimator within each block, which is dictated by the shrinkage factor, is of great importance.

In a closely related setting, though concerning a direct rather than an inverse regression problem involving orthogonal regressors and correlated Gaussian noise, \cite{GoldenshlugerT2001} showed that the standard James-Stein estimator still has theoretically near-optimal risk performance (in comparison with the linear oracle estimator) as long as the correlation is mild. 
In our context, empirical analyses show that, for the sampling design we consider, each block of the noise covariance matrix $\mathbf{V}$ is quite well-conditioned even for higher SH level $l$ (condition number is close to $1$ for $l=4$ and less than 1.2 for $l=10$), while the maximum absolute correlation among the coefficients within each block is modest (around $0.3$). 
Thus, our choice of the shrinkage factor can be seen as a hybrid  addressing the combined scenarios dealt with by \cite{CavalierT2002} and \cite{GoldenshlugerT2001}. This explains the satisfactory empirical  behavior of our proposed estimator.


\section{Synthetic Experiments}\label{sec:simul}

In this section, we first compare the running times of the three FOD estimators, namely, \textbf{BJS}, \textbf{SHridge} and \textbf{SCSD}. Then their performances are assessed 
 through extensive synthetic experiments under different settings in terms of  the number of fibers, separation angles between pairs of fiber bundles, the number of gradient directions, $b$-value, and signal-to-noise ratio (SNR).

\subsection{Running Time Comparison}
\begin{figure}[H]
  \centering
  \includegraphics[width=\textwidth]{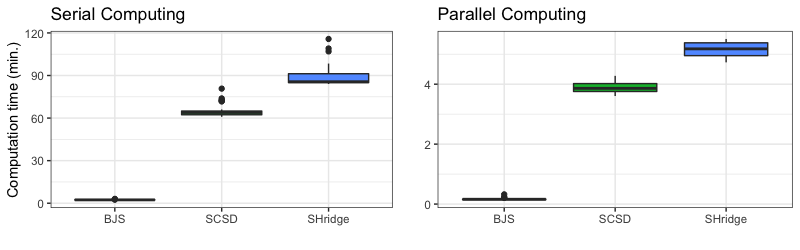}
  \caption{\textbf{Execution times:}  across 30 replicates, applied to 100K voxels with $n=91, l_{\max} = l_{\max}^s = 10$, evaluated on a server with Xeon 72 core processor, 2.3GHz, 256GB RAM.  Left panel -- serial computing. Right panel -- parallel computing with 30 cores. }
  \label{fig:comp_time}
\end{figure}
The execution times of the three methods across $30$ simulation replicates are shown in Fig. \ref{fig:comp_time}. For \textbf{SHridge}, a grid with 100 values is used for tuning parameter selection (with BIC). For \textbf{SCSD},  the additional time by conducting the \textit{superCSD} procedure is reported. 
On average it took \textbf{BJS} 7.246 minutes to process 100K voxels with  $n=91$ gradient directions and  $l_{\max} = l_{\max}^s = 10$ in serial computing on a server with Xeon 72 core processor, 2.3GHz, 256GB RAM.  Moreover, 
Fig. \ref{fig:comp_time} shows that \textbf{BJS} is at least 10 times faster than the other two methods in terms of both serial and parallel  computing. 


\subsection{Experimental Settings}
We consider two fibers crossing at 5 different \textit{separation angles} ($30^{\circ}, 45^{\circ}, 60^{\circ}, 75^{\circ}, 90^{\circ}$), and three fibers crossing at 2 different pair-wise separation angles ($60^{\circ}, 90^{\circ}$). Also, two different $b$-values $1,000s/mm^2,~3,000s/mm^2$, a key parameter in D-MRI experiments, are considered. Moreover, diffusion signals are sampled along, $n = 41, 91, 321$, respectively, gradient directions pointing to the centers of the upper-half triangles  of  an \textit{icosphere mesh} with increasing orders. Lastly, two different levels of SNR ($20, 50$), are used for data generation. Here, SNR is defined as the ratio of the non-diffusion weighted signal intensity $S_0$ to the standard deviation $\sigma$ of the Rician noise. For each setting, 100 independent replicates of diffusion weighted measurements 
are generated by adding the Rician noise to the noiseless diffusion measurements. 

The synthetic experiments cover settings commonly used in both clinical and research purpose D-MRI experiments. Especially, in the HPC application, we have $90$ gradient directions at $b$-value $3,000s/mm^2$ and a median  SNR around $50$ (Fig. \ref{fig:snr}). 

\subsection{Evaluation Metrics}
Estimated FODs are visualized and directly compared with the true fiber directions in Figures \ref{fig:simul_degree45}, \ref{fig:simul_degree30}, \ref{fig:simul_degree607590}, \ref{fig:simul_fib3},  where the opaque color represents the mean of the estimated FOD (across 100 replicates), the semitranslucent color represents the mean plus two standard deviations of the estimated FODs, and the solid lines represent true fiber directions.  
	
 In the HCP application, the purpose of FOD estimation is to obtain fiber direction estimate at each voxel, which is then used as inputs in the tractography algorithm to reconstruct SLF.   Specifically, we use a \textit{peak detection algorithm} \citep{YanCPP2018}  to extract the peak(s) of the estimated FOD  and use the peak direction(s) as the estimated fiber direction(s). 
	The performance in terms of fiber direction estimation is evaluated by three metrics (Tables \ref{table:simul_degree45}, \ref{table:simul_degree30}, \ref{table:simul_degree607590}, \ref{table:simul_fib3}): (i) \textbf{D.R.} --  correct peak detection rate, defined as the percentage of replicates where the peak detection algorithm finds the correct number of fibers; 
	(ii) \textbf{Bias.Sep} -- bias in separation angle estimation, defined as the difference between \textbf{Mean.Sep} and the true separation angle, where \textbf{Mean.Sep} is the  acute separation angle between a pair of estimated fiber directions averaged across those replicates in which the correct number of fibers is detected; 
	and (iii) \textbf{RMSAE},  defined as  the squared-root of mean squared angular error -- the acute angular difference between the true fiber direction and  the corresponding estimated fiber direction -- calculated on those replicates in which the correct number of fibers is detected. 	
Since all angles are acute angles (measured in arc degrees),  we use the usual summary statistics instead of  spherical summaries.  These metrics are reported in Tables \ref{table:simul_degree45}, and \ref{table:simul_degree30}, \ref{table:simul_degree607590}, \ref{table:simul_fib3}. 

\subsection{Results}

\begin{figure}[H]
  \centering
  \includegraphics[width=\textwidth]{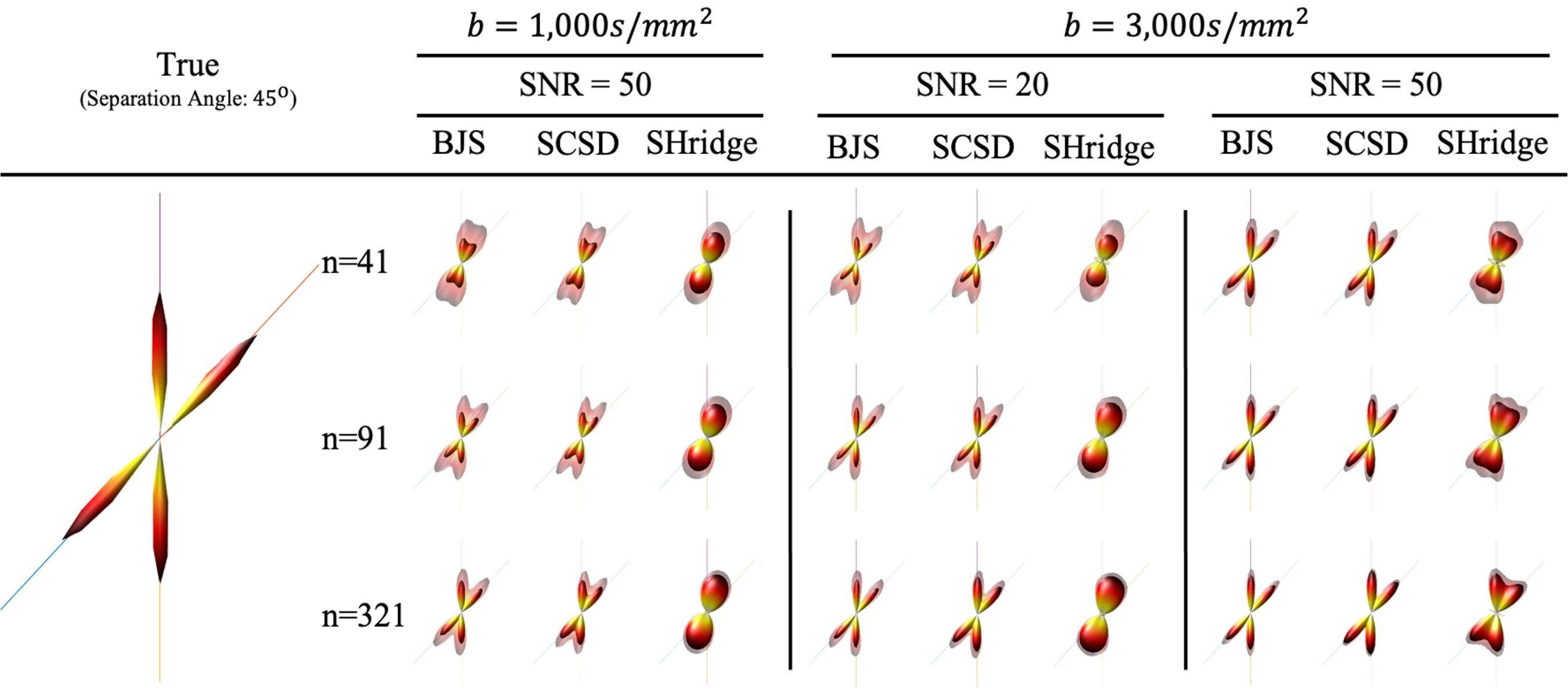}
  \caption{\textbf{Synthetic experiment: two fibers crossing at  $45^{\circ}$.} The solid lines are the true fiber directions. The opaque part and the semitranslucent part represent the mean estimated FODs across 100 replicates and the mean plus two standard deviations of the estimated FODs.}
  \label{fig:simul_degree45}
\end{figure}

In the case of two fibers crossing at a $45^{\circ}$  separation angle  (Fig. \ref{fig:simul_degree45} and Table \ref{table:simul_degree45}),  visually, \textbf{BJS} is the best estimator among the three, as (on average) it shows  the most accurate direction and retains the angular resolution the best.  \textbf{SHridge} performs the worst and shows very poor performance except for  $b = 3,000s/mm^2$, SNR = 50 and $n = 321$. 
Under  $b = 1,000s/mm^2$,  \textbf{SCSD} performs the best in terms of peak detection rate, whereas under $b = 3,000s/mm^2$, both \textbf{BJS} and \textbf{SCSD} could successfully identify two fibers with high rates. In terms of  \textbf{RMSAE}, \textbf{BJS} and \textbf{SCSD} have similar performance and both are much better than \textbf{SHridge}. However,  \textbf{BJS}  has considerably less bias in separation angle estimation than \textbf{SCSD}.  This phenomenon is observed across nearly all simulation settings considered in this section.

In the case of two fibers crossing at  $30^{\circ}$ and $b$-value $3,000s/mm^2$,  \textbf{BJS} outperforms both \textbf{SCSD} and \textbf{SHridge} (Fig. \ref{fig:simul_degree30} and Table  \ref{table:simul_degree30}). In terms of peak detection rates, the performance of \textbf{BJS} is almost twice as good as \textbf{SCSD} and is much better than \textbf{SHridge}. 
Moreover,  \textbf{BJS} has little bias in separation angle estimation, whereas \textbf{SCSD} tends to severely underestimate the separation angle under the high SNR (i.e., 50) settings.

In the case of two fibers crossing at  moderate to large separation angles  ($60^{\circ}, 75^{\circ}, 90^{\circ}$) (Fig. \ref{fig:simul_degree607590} and Table \ref{table:simul_degree607590}), and three fibers crossing at  a pairwise separation angle $60^{\circ}$ or $90^{\circ}$ (Fig. \ref{fig:simul_fib3} and Table \ref{table:simul_fib3}), \textbf{BJS} outperforms   \textbf{SCSD} in the two-fiber cases and has comparable performance with \textbf{SCSD} in the three-fiber cases. Both perform  better than \textbf{SHridge}. 

In summary, when the separation angle is small (i.e., the most challenging settings), \textbf{BJS} has a distinct advantage over \textbf{SCSD} and \textbf{SHridge}. Overall,  \textbf{BJS} performs the best in  separation angle estimation and shows competitive performance in peak detection and fiber direction estimation.

\begin{table}[h]
	\caption{\textbf{Synthetic experiment: two fibers crossing at $45^{\circ}$.} \textbf{D.R.}:  correct peak detection rate;  \textbf{Bias.Sep} (\textbf{s.e.}):   bias (standard error) of separation angle estimation (in arc degree); \textbf{RMSAE}: root mean squared acute angular error (in arc degree) of fiber direction estimation. }
	\label{table:simul_degree45}
	\centering
	\resizebox{\columnwidth}{!}{%
		\begin{tabular}{ccccccccccc}
			\multicolumn{11}{l}{$b=1000s/mm^2, l_{\max}^s=12$}                                                                                                                                                                                                                                                                                                                                                                                                                                           \\ \hline
			\multicolumn{1}{|c|}{\multirow{2}{*}{\textbf{SNR}}} & \multicolumn{1}{c|}{\multirow{2}{*}{\textbf{Design}}} & \multicolumn{3}{c|}{\textbf{BJS}}                                                                                        & \multicolumn{3}{c|}{\textbf{SCSD}}                                                                                       & \multicolumn{3}{c|}{\textbf{SHridge}}                                                                                    \\ \cline{3-11} 
			\multicolumn{1}{|c|}{}                              & \multicolumn{1}{c|}{}                                 & \multicolumn{1}{c|}{\textbf{D.R.}} & \multicolumn{1}{c|}{\textbf{Bias.Sep (s.e.)}} & \multicolumn{1}{c|}{\textbf{RMSAE}} & \multicolumn{1}{c|}{\textbf{D.R.}} & \multicolumn{1}{c|}{\textbf{Bias.Sep (s.e.)}} & \multicolumn{1}{c|}{\textbf{RMSAE}} & \multicolumn{1}{c|}{\textbf{D.R.}} & \multicolumn{1}{c|}{\textbf{Bias.Sep (s.e.)}} & \multicolumn{1}{c|}{\textbf{RMSAE}} \\ \hline
			\multicolumn{1}{|c|}{\multirow{3}{*}{\textbf{50}}}  & \multicolumn{1}{c|}{$n= 41, l_{\max}=6$}              & \multicolumn{1}{c|}{62\%}          & \multicolumn{1}{c|}{-1.33 (0.78)}             & \multicolumn{1}{c|}{10.89}          & \multicolumn{1}{c|}{82\%}          & \multicolumn{1}{c|}{-4.90 (0.38)}             & \multicolumn{1}{c|}{10.20}          & \multicolumn{1}{c|}{3\%}           & \multicolumn{1}{c|}{29.37 (8.95)}             & \multicolumn{1}{c|}{71.91}          \\ \cline{2-11} 
			\multicolumn{1}{|c|}{}                              & \multicolumn{1}{c|}{$n= 91, l_{\max}=10$}             & \multicolumn{1}{c|}{83\%}          & \multicolumn{1}{c|}{-0.59 (0.59)}             & \multicolumn{1}{c|}{10.37}          & \multicolumn{1}{c|}{97\%}          & \multicolumn{1}{c|}{-4.60 (0.23)}             & \multicolumn{1}{c|}{9.51}           & \multicolumn{1}{c|}{0\%}           & \multicolumn{1}{c|}{-}                        & \multicolumn{1}{c|}{-}              \\ \cline{2-11} 
			\multicolumn{1}{|c|}{}                              & \multicolumn{1}{c|}{$n= 321, l_{\max}=12$}            & \multicolumn{1}{c|}{100\%}         & \multicolumn{1}{c|}{-0.16 (0.31)}             & \multicolumn{1}{c|}{6.13}           & \multicolumn{1}{c|}{100\%}         & \multicolumn{1}{c|}{-4.87 (0.21)}             & \multicolumn{1}{c|}{6.02}           & \multicolumn{1}{c|}{0\%}           & \multicolumn{1}{c|}{-}                        & \multicolumn{1}{c|}{-}              \\ \hline
			\multicolumn{11}{l}{}                                                                                                                                                                                                                                                                                                                                                                                                                                                                        \\
			\multicolumn{11}{l}{$b=3000s/mm^2, l_{\max}^s=12$}                                                                                                                                                                                                                                                                                                                                                                                                                                           \\ \hline
			\multicolumn{1}{|c|}{\multirow{2}{*}{\textbf{SNR}}} & \multicolumn{1}{c|}{\multirow{2}{*}{\textbf{Design}}} & \multicolumn{3}{c|}{\textbf{BJS}}                                                                                        & \multicolumn{3}{c|}{\textbf{SCSD}}                                                                                       & \multicolumn{3}{c|}{\textbf{SHridge}}                                                                                    \\ \cline{3-11} 
			\multicolumn{1}{|c|}{}                              & \multicolumn{1}{c|}{}                                 & \multicolumn{1}{c|}{\textbf{D.R.}} & \multicolumn{1}{c|}{\textbf{Bias.Sep (s.e.)}} & \multicolumn{1}{c|}{\textbf{RMSAE}} & \multicolumn{1}{c|}{\textbf{D.R.}} & \multicolumn{1}{c|}{\textbf{Bias.Sep (s.e.)}} & \multicolumn{1}{c|}{\textbf{RMSAE}} & \multicolumn{1}{c|}{\textbf{D.R.}} & \multicolumn{1}{c|}{\textbf{Bias.Sep (s.e.)}} & \multicolumn{1}{c|}{\textbf{RMSAE}} \\ \hline
			\multicolumn{1}{|c|}{\multirow{3}{*}{\textbf{20}}}  & \multicolumn{1}{c|}{$n= 41, l_{\max}=6$}              & \multicolumn{1}{c|}{92\%}          & \multicolumn{1}{c|}{-1.71 (0.61)}             & \multicolumn{1}{c|}{11.4}           & \multicolumn{1}{c|}{88\%}          & \multicolumn{1}{c|}{-3.39 (0.52)}             & \multicolumn{1}{c|}{8.77}           & \multicolumn{1}{c|}{2\%}           & \multicolumn{1}{c|}{21.65 (16.05)}            & \multicolumn{1}{c|}{63.43}          \\ \cline{2-11} 
			\multicolumn{1}{|c|}{}                              & \multicolumn{1}{c|}{$n= 91, l_{\max}=10$}             & \multicolumn{1}{c|}{97\%}          & \multicolumn{1}{c|}{-1.67 (0.31)}             & \multicolumn{1}{c|}{7.79}           & \multicolumn{1}{c|}{98\%}          & \multicolumn{1}{c|}{-2.59 (0.28)}             & \multicolumn{1}{c|}{6.76}           & \multicolumn{1}{c|}{0\%}           & \multicolumn{1}{c|}{-}                        & \multicolumn{1}{c|}{-}              \\ \cline{2-11} 
			\multicolumn{1}{|c|}{}                              & \multicolumn{1}{c|}{$n= 321, l_{\max}=12$}            & \multicolumn{1}{c|}{100\%}         & \multicolumn{1}{c|}{-1.99 (0.22)}             & \multicolumn{1}{c|}{5.73}           & \multicolumn{1}{c|}{100\%}         & \multicolumn{1}{c|}{-1.61 (0.20)}             & \multicolumn{1}{c|}{3.84}           & \multicolumn{1}{c|}{1\%}           & \multicolumn{1}{c|}{-2.08 (-)}                & \multicolumn{1}{c|}{5.50}           \\ \hline
			\multicolumn{1}{|c|}{\multirow{3}{*}{\textbf{50}}}  & \multicolumn{1}{c|}{$n= 41, l_{\max}=6$}              & \multicolumn{1}{c|}{100\%}         & \multicolumn{1}{c|}{-1.03 (0.29)}             & \multicolumn{1}{c|}{5.10}           & \multicolumn{1}{c|}{100\%}         & \multicolumn{1}{c|}{-4.55 (0.22)}             & \multicolumn{1}{c|}{5.27}           & \multicolumn{1}{c|}{27\%}          & \multicolumn{1}{c|}{6.28 (1.06)}              & \multicolumn{1}{c|}{9.45}           \\ \cline{2-11} 
			\multicolumn{1}{|c|}{}                              & \multicolumn{1}{c|}{$n= 91, l_{\max}=10$}             & \multicolumn{1}{c|}{98\%}          & \multicolumn{1}{c|}{-0.05 (0.20)}             & \multicolumn{1}{c|}{3.21}           & \multicolumn{1}{c|}{100\%}         & \multicolumn{1}{c|}{-2.77 (0.16)}             & \multicolumn{1}{c|}{3.90}           & \multicolumn{1}{c|}{33\%}          & \multicolumn{1}{c|}{0.24 (1.37)}              & \multicolumn{1}{c|}{7.33}           \\ \cline{2-11} 
			\multicolumn{1}{|c|}{}                              & \multicolumn{1}{c|}{$n= 321, l_{\max}=12$}            & \multicolumn{1}{c|}{100\%}         & \multicolumn{1}{c|}{-0.34 (0.17)}             & \multicolumn{1}{c|}{2.29}           & \multicolumn{1}{c|}{100\%}         & \multicolumn{1}{c|}{-0.77 (0.18)}             & \multicolumn{1}{c|}{2.52}           & \multicolumn{1}{c|}{99\%}          & \multicolumn{1}{c|}{2.42 (0.36)}              & \multicolumn{1}{c|}{4.41}           \\ \hline
			\end{tabular}%
	}
	
\end{table}

\section{HCP Young-Adult Application}
\label{sec:application}
In this section, we investigate the association of SLF lateralization with gender and handedness using HCP young-adult data. 
The WU-Minn Human Connectome Project  (HCP) \citep{VANESSEN2013} 
has eddy-current-corrected 3T D-MRI data of 1206 healthy young adults (Age: 22 $\sim$ 35) from 457 unique families. D-MRI are taken at 3 different \textit{b}-values ($1,000s/mm^2, 2,000s/mm^2, 3,000s/mm^2$) on a $145\times 174\times 145$ grid with voxel size $1.25\times 1.25\times 1.25 mm^3$. For each \textit{b}-value, 90 gradient directions and 6 $b_0$ images are available. In the subsequent analysis, D-MRI measurements with  \textit{b}-value$=3,000s/mm^2$ are used. 

\begin{figure}[H]
	\centering
	\includegraphics[width=0.6\textwidth]{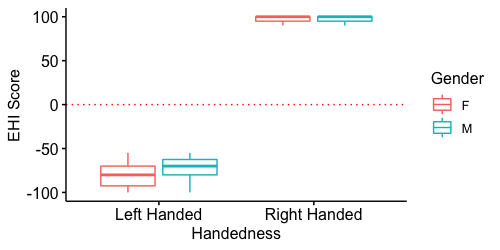}
	\caption{\textbf{EHI distribution by gender-handedness group of the $184$ sampled HCP subjects}}
	\label{fig:handedness}
\end{figure}

We classify the subjects to be left-handed (EHI: -100 $\sim$ -55) and right-handed (EHI: 85 $\sim$ 100) according to the \textit{Edinburgh Handedness Index (EHI)}. In order to remove family effects, we choose at most one subject from each family. If all subjects from a family are right handed, then a subject is randomly selected. Otherwise, priority is given to left handed members.  We also balance the sample in terms of gender by  \textit{stratified sampling} according to the EHI. Through the above sampling scheme, 184 subjects (left-handed female: 23; left-handed male: 23; right-handed female: 69; right-handed male: 69)  are selected. The EHI distribution of these $184$ subjects by the gender-handedness group is shown in Fig. \ref{fig:handedness}.

Our data analysis pipeline is shown in Fig. \ref{fig:pipeline} and the details are described as follows. 

\subsection{Preprocessing}

D-MRI data downloaded from the HCP database -- \textit{ConnectomeDB} have gone through basic quality control (QC)  and have also been minimally preprocessed. Detailed information about the \textit{HCP minimal preprocessing pipeline} can be found in \cite{GLASSER2013}. The steps include: (i) Intensity normalization; (ii) EPI distortion correction; (iii) Eddy current correction; (iv) Gradient nonlinearity correction; (v) Registration  of the mean $b_0$ image (T2w image) to the native volume T1w image; and the transformation of diffusion data, gradient deviation, and the gradient directions to the \textit{structural space (T1w space)}. Thus, the HCP D-MRI data have already been \textit{co-registered} to the structural space.

We perform additional processing on each D-MRI image using the software \textit{FSL} version 6.0.0 \citep{JENKINSON2012} and R packages  \textit{fslr} \citep{fslr} and \textit{neurohcp} \citep{neurohcp} from the \textit{neuroconductor} repository.
The original T1w image contains both skull and the brain. Since HCP D-MRI data have already been co-registered to the structural (T1w) space, we apply the T2w extracted binary brain mask provided by HCP onto the original T1w image to  obtain the T1w extracted brain image. Using the T1w extracted brain image and  the \textit{FAST} segmentation algorithm \citep{Zhang2001} in \textit{FSL}, each voxel in the brain is classified into three different tissue types (CSF – cerebrospinal fluid, GM – grey matter, WM – white matter). The segmentation result is used to create a \textit{white-matter mask}. Hereafter, we refer to voxels within the white-matter mask as the \textit{white-matter voxels}. 
Moreover,  the T1w image is registered to a standard space -- \textit{MNI152\_T1\_2mm}(\url{http://www.bic.mni.mcgill.ca/ServicesAtlases/HomePage}) -- by the \textit{FSL} registration tools  FLIRT \citep{JENKINSON2002825} (for initial linear registration) and  FNIRT \citep{WOOLRICH2009S173} (for subsequent nonlinear registration).

\subsection{SLF Masks}
For 	\textit{Superior Longitudinal Fasciculus (SLF)}  reconstruction, we adopt a regional-seeding tractography strategy. See Section \ref{sec:tractography} for a detailed discussion. For this purpose, we need to create region of interest (ROI) masks that contain SLF.

\begin{figure}[H]
	\centering
	\includegraphics[width=\textwidth]{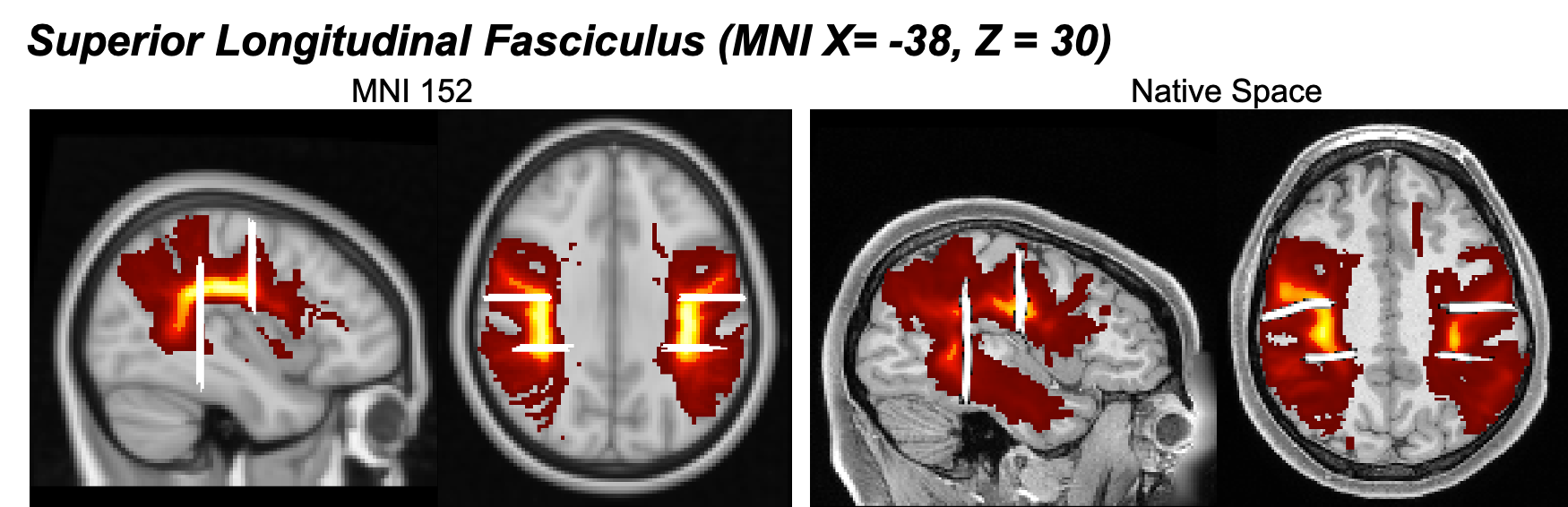}
	\caption{\textbf{Left Panel -- SLF masks on template space (left), Right Panel -- SLF masks on the native space of one HCP subject:} The (probabilistic) ROI masks are shown by the heatmap, where brighter color corresponds to a higher probability. The (binary) streamline selection masks are shown by the white-colored strips. 
	}
	\label{fig:registration}
\end{figure}

On the MNI152\_T1 template space, we use \textit{FSLeyes} \citep{fsleyes} and the \textit{JHU White-Matter Tractography Atlas} \citep{WAKANA2007,HUA2008} to create the SLF masks in left- and right- hemispheres, respectively. The left- SLF ROI contains 41,694 voxels and the right- SLF ROI contains 38,386 voxels. 
Since it is known that SLF and the corticospinal tract (CST) are crossing, we further use binary masks from \textit{AutoPtx} \citep{autoptx} for streamline selection to dissect SLF from the initial tractography results. As can be seen on Fig. \ref{fig:registration}, the binary masks are situated at the margins of the portions of SLF masks where the probability of being on SLF is high (indicated by bright color).

Since all subsequent analyses are conducted on the subject native space, we use the inverse transformation (derived from the registration step) to move masks on the template space back to the subject native space. On the subject native space, the numbers of voxels in left- SLF and right- SLF ROIs are $47,379\pm5,912$ and $42,772\pm5,430$, respectively. More details on  preprocessing and ROI masks creation can be found on \url{https://github.com/vic-dragon/BJS}.

\subsection{FOD Estimation and Peak Detection}
\label{sec:hcp-fod}
For each subject, we first use the \textit{white-matter voxels} to  estimate the \textit{response function} $R(\cdot)$ in the FOD model (\ref{eq:dMRI_signal}) following \cite{YanCPP2018}. Specifically, at each voxel, we fit the single tensor model (\ref{eq:single-tensor}) and identify voxels with FA value (\ref{eq:FA})   greater than 0.8 and the ratio between the two smaller eigenvalues less than 1.5 as having a single dominant fiber bundle. These voxels are then used to determine the leading and minor eigenvalues of a tensor and the response function is  defined as the diffusion signal under the single tensor model with this tensor. 
 See the Supplementary Text (Section \ref{sec:supp-response}) for more details. 

For each subject, we also estimate the signal-to-noise-ratio (SNR: =$S_0/\sigma$)  using the 6 $b_0$ images and the overall SNR is taken as the median  SNR  over all voxels. The estimated SNR and response function of the $184$ selected HCP subjects are shown in Fig. \ref{fig:snr}.

\textbf{BJS} estimators  are then derived for white-matter voxels within the SLF masks.  The peak directions of the estimated FODs are extracted by a \textit{peak detection algorithm} \citep{YanCPP2018}. Moreover, non-\textit{white-matter voxels} within SLF  masks are automatically specified as isotropic and thus have no associated peak direction. The peak detection algorithm associates each voxel with either none, one or multiple directions and these are used as inputs in a deterministic tracking algorithm \textit{DiST} \citep{WongLPP2016} for SLF reconstruction described in Section \ref{sec:tractography}.

\subsection{SLF Reconstruction by Tractography and Streamline Selection}
\label{sec:tractography}

In neuroscience, tractography refers to the technique of reconstructing and visually representing white matter fibers using D-MRI data. While applying a tractography algorithm, there are several options for \textit{seeding and terminating criteria}. Tracking is initialized at so-called seed locations and there are generally two options: \textit{whole-brain seeding} vs. seeding within a \textit{region of interest (ROI)}, referred to as \textit{regional-seeding}. Commonly used terminating criteria include trajectory bending more than a prespecified angle in a single step; trajectory entering a region of low FA or leaving the white-matter segment. Moreover, tracking may be terminated when the trajectory leaves the ROI. For an overview of deterministic tractography, see \cite{Alexander2010}.

Here we apply the \textit{DiST} tracking algorithm  (\url{https://github.com/vic-dragon/dmri.tracking}),  a deterministic tractography algorithm that can handle zero or multiple directions within one voxel and thus is suitable for tracking in crossing fiber regions  \citep{WongLPP2016}. Moreover, we use the probabilistic masks for SLF (one on each hemisphere) from the \textit{JHU White-Matter Tractography Atlas} \citep{WAKANA2007,HUA2008} as both the seeding mask and the terminating mask, meaning that tracking starts from every (white-matter) voxel within these masks and trajectories will be terminated while leaving the SLF region specified by these masks. Another stopping criterion we use is when there is no viable voxel within two steps, where non-viable voxels are those leading to trajectory bending more than 60 degrees or being isotropic (e.g., non-white-matter voxels).

Note that SLF crosses with other fiber tracts, mainly, with the corticospinal tract (CST). As can be seen from the orientation color map of one HCP subject (left panel of Fig. \ref{fig:colormap}), the SLF region crosses with CST (indicated by blue color as this tract is mainly along the inferior-superior direction). As a result, the reconstructed fibers contain not only those of SLF, but also some of CST. This can also be seen from the tractography results of one HCP subject (right panel of Fig. \ref{fig:colormap}), which shows a big bundle of blue-colored tracks. In order to better dissect SLF, we further apply \textit{streamline selection}. Here, we use binary masks from \textit{AutoPtx} to dissect SLF from the initial tractography results. Only tracks (streamlines) that pass through both \textit{AutoPtx} binary masks are retained. 

The above regional-seeding approach is suitable for extracting a specific pathway (here SLF) or mapping tracts from a specific region. One advantage of the regional-seeding approach to the whole-brain-seeding approach is that the former is computationally much less intensive and scales better to processing a large number of subjects/images. The regional-seeding approach may also take advantage of existing knowledge in brain anatomy. A potential disadvantage of a regional-seeding approach is that it may lead to incomplete tract reconstruction. This can be mitigated by using anatomically informed masks such as those from a white matter atlas as we have done here. In Fig. \ref{fig:slf_seeding}, we show SLF reconstruction results of one HCP subject after streamline selection by the \textit{AutoPtx} masks under different seeding strategies. 
It shows that the regional-seeding approach we adopted here does not lose too many fiber tracks compared to the whole-brain-seeding approach.

\subsection{Feature Extraction}
After tractography and streamline selection, brain structural connectivity features can be extracted including  the number of streamlines, the length of streamlines, etc. Here we focus on the difference between the left- and right- hemispheric SLF for the purpose of investigating the lateralization pattern of SLF and its association with gender and handedness.

Specifically, for each subject, we calculate a \textit{lateralization score (LS)} based on the relative difference between the numbers of selected streamlines from the left- and right- hemispheric SLF, respectively:
\begin{equation}
	LS = \frac{\mbox{Streamlines in Left SLF - Streamlines in Right SLF}}{\mbox{(Streamlines in Left SLF + Streamlines in Right SLF)/2}}
\end{equation}

Here, the denominator serves the purpose of normalization so that the LS from subjects with different brain sizes are comparable. As can be seen from Fig \ref{fig:scatter}, the LS is uncorrelated with the size of the SLF ROI. A similar score was used by \cite{Catani2007} to quantify lateralization of the language pathway.

\subsection{Group Analysis and Results}
\begin{figure}[H]
  \centering
  \includegraphics[width=\textwidth]{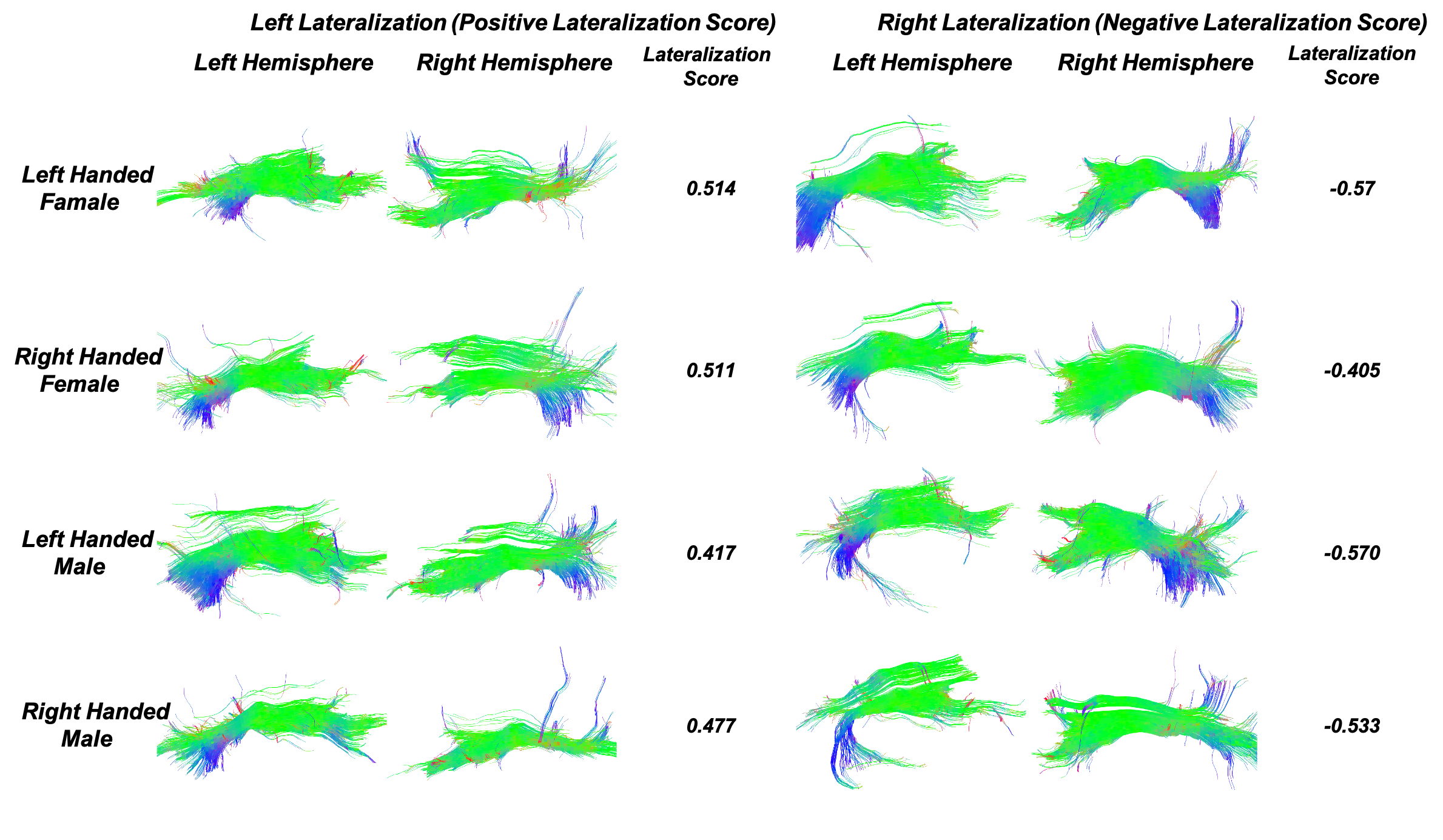}
  \caption{\textbf{Reconstructed  SLF of representative HCP subjects.}  (color scheme: green: anterior-posterior; blue: superior-inferior; red: left-right)}
  \label{fig:tract3}
\end{figure}
\begin{figure}[H]
	\centering
	\includegraphics[width=\textwidth]{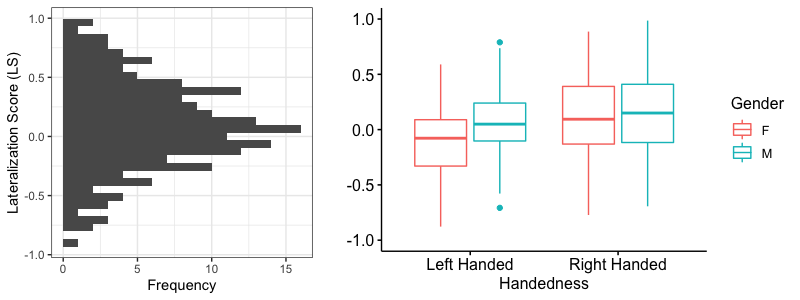}
	\caption{\textbf{Lateralization score distribution:} Left panel -- on  all subjects; Right panel -- by gender-handedness group}
	\label{fig:lateralization_score}
\end{figure}

Reconstructed SLF  of representative subjects from each gender-handedness group with positive- and negative- lateralization scores are illustrated in Fig. \ref{fig:tract3}.  Moreover, the lateralization score distribution by the gender-handedness group is shown in Fig. \ref{fig:lateralization_score}.

We use a \textit{two-way ANOVA} model to study the association between the SLF lateralization score and gender and handedness:
\[
Y_{ijk} = \mu + \alpha_i + \beta_j + \gamma_{ij} + \varepsilon_{ijk},  ~~i=1,2, ~ j=1,2, ~ k=1,\cdots, n_{ij},
\]
where $Y_{ijk}$ is the lateralization score, $\mu$ is the overall mean, $\alpha_i$ is the main effect of handedness at level $i$ ($i=1$: left-handed, $i=2$: right-handed), $\beta_j$ is the main effect of gender at level $j$ ($j=1$: female, $j=2$: male), $\gamma_{ij}$ is the interaction effect between handedness and gender for the level combination $i, j$, with constraints $\sum_{i=1}^2 \alpha_i = 0, \sum_{j=1}^2 \beta_i = 0, \sum_{i=1}^2 \gamma_{ij} = 0, \sum_{j=1}^2 \gamma_{ij} = 0 $, and  $\varepsilon_{ijk}$ are i.i.d $N(0,\sigma^2)$ errors. The diagnostic plots (Fig. \ref{fig:diagnostic})  show a good model fit. 

\begin{table}[H]
\caption{\textbf{HCP D-MRI application: two-way ANOVA table}}
\label{table:twowayanova}
\begin{tabular}{|c|c|c|c|c|c|}
\hline
\textbf{}                    & \textbf{d.f.} & \textbf{SS} & \textbf{MS} & \textbf{F-value} & \textbf{$p$-value} \\ \hline
\textbf{Handedness}          & 1             & 0.937       & 0.937       & 6.742            & 0.0102             \\ \hline
\textbf{Gender}              & 1             & 0.226       & 0.226       & 1.629            & 0.2035             \\ \hline
\textbf{Handedness * Gender} & 1             & 0.265       & 0.265       & 1.907            & 0.1690             \\ \hline
\textbf{Residuals}           & 180           & 25.016      & 0.139       &                  &                    \\ \hline
\end{tabular}
\end{table}
According to the two-way ANOVA Table (Table \ref{table:twowayanova}), SLF lateralization is significantly associated with handedness. Moreover, the 95$\%$ confidence interval of the contrast between left-handedness and right-handedness is $(-0.289, -0.041)$, which suggests right-handed subjects have a greater left lateralization tendency in SLF (i.e., larger LS) compared to left-handed subjects. On the other hand, there is no significant gender effect or gender-handedness interaction effect on SLF lateralization score. 
 
 \subsection{HCP D-MRI Application with DSI Studio}
We also conducted the HCP application using \textit{DSI Studio}  (\url{http://dsi-studio.labsolver.org/}) -- a tractography software tool for D-MRI analysis \citep{YehWT2010,YehVW2013}. \textit{DSI Studio} uses \textit{Orientation Distribution Function (ODF)} as a local fiber estimation method. 
ODF  is the projection of the diffusion probability density onto the surface of the unit sphere along a ray 
emanating from a voxel center \citep{Tuch02, Tuch04}. One limitation of ODF is that it does not preserve the sharp features associated with the underlying fiber bundle orientation.  Therefore,  if the objective is 
white matter fiber tracts reconstruction, it is expected that the FOD model is more efficient  as it directly models the distribution of fiber orientation  
within a voxel.  

By conducting the HCP application using \textit{DSI Studio}, we obtain qualitatively similar but less significant results (Fig.,  \ref{fig:lateralization_score_dsi}; Table  \ref{table:twowayanova_dsi}). Specifically, the p-value associated with the handedness effect is $0.0599$ (vs. $0.0102$ from our data analysis pipeline), whereas there is also no significant gender or handedness-gender interaction effect. More details can be found in 
the Supplementary Text (Section \ref{sec:supp-dsi}). 

\section{Discussion}
\label{sec:discussion}

In this paper, we investigate the association between brain structural connectivity and demographic and behavioral features using D-MRI data from the Human Connectome Project (HCP).  
Specifically,  we derive a  lateralization score for a major association tract,	\textit{Superior Longitudinal Fasciculus (SLF)}, 
and relate it to gender and handedness. We find significant handedness effects, indicating a difference in SLF lateralization between left-handed and right-handed individuals. 
Moreover, we propose a novel computationally efficient method, \textbf{BJS}, for 
estimating the fiber orientation distribution (FOD) at each brain voxel. We also  establish a D-MRI data analysis pipeline that could be utilized for population level associative studies for relating brain anatomic features to  external features including demographic, behavioral, or cognitive measurements.

The proposed \textbf{BJS} method is scalable for statistical analysis of brain structural connectivity at a population level. 
The \textbf{BJS} procedure constitutes an effective improvisation of the classical James-Stein shrinkage that solves an ill-conditioned problem
with noisy measurements in a non-standard setting where an exact diagonalization of the convolution operator in a basis representing the observation vector (D-MRI measurements) is not feasible due to finite sampling effects.

To reconstruct white matter fiber tracts, the estimated FOD at each voxel needs to  provide reasonably accurate fiber orientation information. Based on synthetic experiment results, we believe that the estimated FODs via \textbf{BJS} are accurate enough to be used as inputs to a tracking algorithm.
Although a large proportion of the neuronal fiber bundles can be explained by the reconstructed neuronal fiber tracks based on D-MRI, it is not sufficient to represent the actual fiber system in the brain. Also, the estimated fiber composition can be different depending on the tractography algorithm \citep{jones2013}. Despite these challenges, this paper demonstrates that it is possible to extract meaningful structural connectivity information from reconstructed neuronal fiber tracts based on D-MRI data and to relate such information with external features. 

%
%

\section*{Acknowledgments}
The authors would like to thank Hao Yan, Jilei Yang, and Raymond Wong for sharing codes for peak detection, fiber tracking and generating synthetic D-MRI data. Data used in the real application were provided by the Human Connectome Project, WU-Minn Consortium (Principal Investigators: David Van Essen and Kamil Ugurbil; 1U54MH091657) funded by the 16 NIH Institutes and Centers that support the NIH Blueprint for Neuroscience Research; and by the McDonnell Center for Systems Neuroscience at Washington University. This research is supported by the following grants:  NSF DMS 1713120 (D.P.), DMS-1811405(T.L., D.P.), DMS-1811661(T.L.), DMS-1915894(S.H., D.P., J.P.), DMS-1916125 (S.H., T.L.), and NIH 1R01EB021707(D.P., J.P.).

%
%

\begin{supplement}
	\noindent \textbf{Supplementary Text:} A supplementary text with additional details on FOD estimators,  synthetic experiments results  and the HCP D-MRI application.\\
	\noindent \textbf{Codebase:} Codes and example scripts for synthetic experiments and the HCP application can be found at \url{https://github.com/vic-dragon/BJS}, together with a detailed manual on D-MRI batch downloading and preprocessing. 
\end{supplement}
\bibliographystyle{imsart-nameyear} 
\bibliography{braindraft, dMRI_review}       


\clearpage
\newpage

	\setcounter{page}{1}
	\setcounter{section}{0}
	\renewcommand{\thesection}{S.\arabic{section}}
	\setcounter{subsection}{0}
	\renewcommand{\thesubsection}{S.\arabic{section}.\arabic{subsection}}
	\setcounter{equation}{0}
	\renewcommand{\theequation}{S.\arabic{equation}}
	\setcounter{figure}{0}
	\renewcommand{\thefigure}{S.\arabic{figure}}
	\setcounter{table}{0}
	\renewcommand{\thetable}{S.\arabic{table}}
	\setcounter{proposition}{0}
	\renewcommand{\theproposition}{S.\arabic{proposition}}
	\setcounter{lemma}{0}
	\renewcommand{\thelemma}{S.\arabic{lemma}}
	\setcounter{corollary}{0}
	\renewcommand{\thecorollary}{S.\arabic{corollary}}
	
	\begin{center}
	{\Huge Supplementary Text\\}
	\end{center}

	\section{FOD Estimators: Additional Details} 
		
		\subsection{superCSD Procedure}
	\label{sec:supp-scsd}
	Consider an $l^s_{\max}$ order SH presentation. Let $L_s$ be the corresponding number of SH basis functions. Let 
	$\bs \Phi^s_{n \times L_s}$ be the evaluation matrix of the $L_s$ SH basis functions evaluated on  the $n$ sampled gradient directions. Let $\bs\Phi^{ds}_{n_d\times L_s}$ be the evaluation matrix of the $L_s$ SH basis functions evaluated on a dense  grid (e.g.,  from an icosphere mesh with $n_d=2562$).  Let $\mbf{R}^s_{L_s \times L_s}$ be the $\mbf{R}$ matrix corresponding to the $l^s_{\max}$ order SH representation of the response function. The \textit{superCSD} procedure is as follows: 
	
	\begin{enumerate}
		\item \textbf{Initial step}: Get an initial estimator $\widehat{\mathbf{f}}_0$ (e.g., by \textbf{SHridge}). 
		\item \textbf{Filter step}: In the $l^s_{\max}$ order SH representation of $\widehat{\mathbf{f}}_0$, set the spherical harmonics  coefficients  over order $l=4$ to zero to reduce high frequency noise.
		
		\item \textbf{$(k+1)$th updating step}: Define
		$$\hat{\mathbf{F}}^k= \bs \Phi^{ds} \hat{\mathbf{f}}^k$$ as the estimated FOD  on the dense evaluation grid from the $k$th step.
		  Let
		
		\begin{equation}\label{eq:super_csd}
			\widehat{\mathbf{f}}^{k+1} = \arg\min_{\mathbf{f}} \parallel
			\mathbf{y} - \mathbf{\Phi}^{s} {\mathbf{R}^s}  \mathbf{f}\parallel_2^2 +
			\lambda\parallel \bs \Phi^{k} \mathbf{f}\parallel_2^2, 
		\end{equation}
		
		where $\bs \Phi^k$ is an $n_d\times L_s$ matrix,   
		
		\begin{equation}\label{eq:super_csd_L}
			\bs \Phi^k_{i,(l,m)} := \begin{cases}
				\bs \Phi^{ds}_{i,(l,m)} & \text{if } \hat{F}^k_{i} \leq \tau  \\
				0      & \text{if } \hat{F}^k_{i} > \tau
			\end{cases}, ~~~i=1,\cdots, n_d, ~~l=0,2,\cdots, l^s_{\max},~~ m=-l,\cdots,0,\cdots, l.
		\end{equation}
	
	 Note that, (\ref{eq:super_csd}) penalizes small (including negative) values of the estimated FOD. 
	 
		\item Repeat step 3 until $\bs \Phi^k$ stabilizes.
	\end{enumerate}
	The recommended values  for $\tau>0, \lambda>0$ in \cite{tournier2007} are $0.1, 1$, respectively. As for $l^s_{\max}$, the results from \cite{tournier2007} suggest relatively small level, e.g.,  $l^s_{\max}=8$ for large separation angles, and relatively large level,  e.g.,  $l^s_{\max}=12$ for small separation angles.

	\subsection{BJS}
	\label{sec:supp-bjs}
	\begin{lemma}\label{lemma:quad_from_tail_bound} \textbf{\citep{laurent2000}}
		Let $(w_1,\ldots,w_{2l+1})$ be i.i.d Gaussian variables, with mean 0 and variance 1 and $\lambda_1,\ldots,\lambda_{2l+1}$ be nonnegative. We set
		\[
		\|\bs\lambda\|_1=\sum_{i=1}^{2l+1}|\lambda_i|, ~~~ \|\bs\lambda\|_2=\sqrt{\sum_{i=1}^{2l+1}\lambda_i^2}, ~~~ \|\bs\lambda\|_{\infty}=\sup_{i=1,...,2l+1}|\lambda_i|
		\]
		Then, the following inequality holds for any positive $t$:
		\[
		P(\sum_{i=1}^{2l+1}\lambda_i w_i^2 \geq \|\bs\lambda\|_1 + 2\|\bs\lambda\|_2\sqrt{t} + 2\|\bs\lambda\|_{\infty}t) \leq \exp(-t)
		\]
	\end{lemma}

\subsubsection*{Post-estimation sharpening.}
	
	Let $\hat{F} = \bs \Phi^d_{n_d \times L}\hat{f}$ be the evaluation of the estimated FOD on a dense spherical grid (e.g., $n_{d} = 2562$). Let $J = \{i:\hat{{F}}_i<0, i=1,...,n_d\}$, $|J| = n_{neg}$. Let $L_{s}$ be the number of SH basis functions corresponding to $l_{\max}^s$ order. 
	Let $\bs\Phi^s_{n\times L_s}$ be the evaluation matrix of the $L_s$ SH basis functions on the $n$ sampled gradient directions. Let $\bs\Phi^{neg}_{n_{neg}\times L_s}$ be the evaluation matrix of the $L_s$ SH basis functions on $n_{neg}$ grid points corresponding to negatively estimated FOD values. Let $\mbf{R}^s_{L_s \times L_s}$ be the $\mbf{R}$ matrix corresponding to the $l^s_{\max}$ order SH representation of the response function. 
	
	Then the (one-step) updated estimates of the SH coefficients of FOD is defined as: 
		\begin{equation}
		\hat{\mbf{f}}^{BJS} = \arg\min_{\mbf{f}}||\tilde{\mbf{y}} - \tilde{\mbf{Z}}\mbf{f}||^2_2,
		\end{equation}
		where
		\[
		\tilde{\mbf{y}}=\begin{bmatrix}
		\mbf{y}\\
		\mbf{0}\\
		\end{bmatrix}, \tilde{\mbf{Z}}_{(n+n_{neg})\times L_s} = \begin{bmatrix}
		\bs\Phi^s\mbf{R}^s\\
		\bs\Phi^{neg}\\
		\end{bmatrix}
		\]
		\[
		\hat{\mbf{f}}^{BJS} = (\tilde{\mbf{Z}}^T \tilde{\mbf{Z}})^{-1}\tilde{\mbf{Z}}^T\tilde{\mbf{y}}
		\]

	\section{Synthetic Experiments: Additional Results} 
	Here we provide additional plots and tables of the synthetic experiments.
	\begin{figure}[h]
		\centering
		\includegraphics[width=0.8\textwidth]{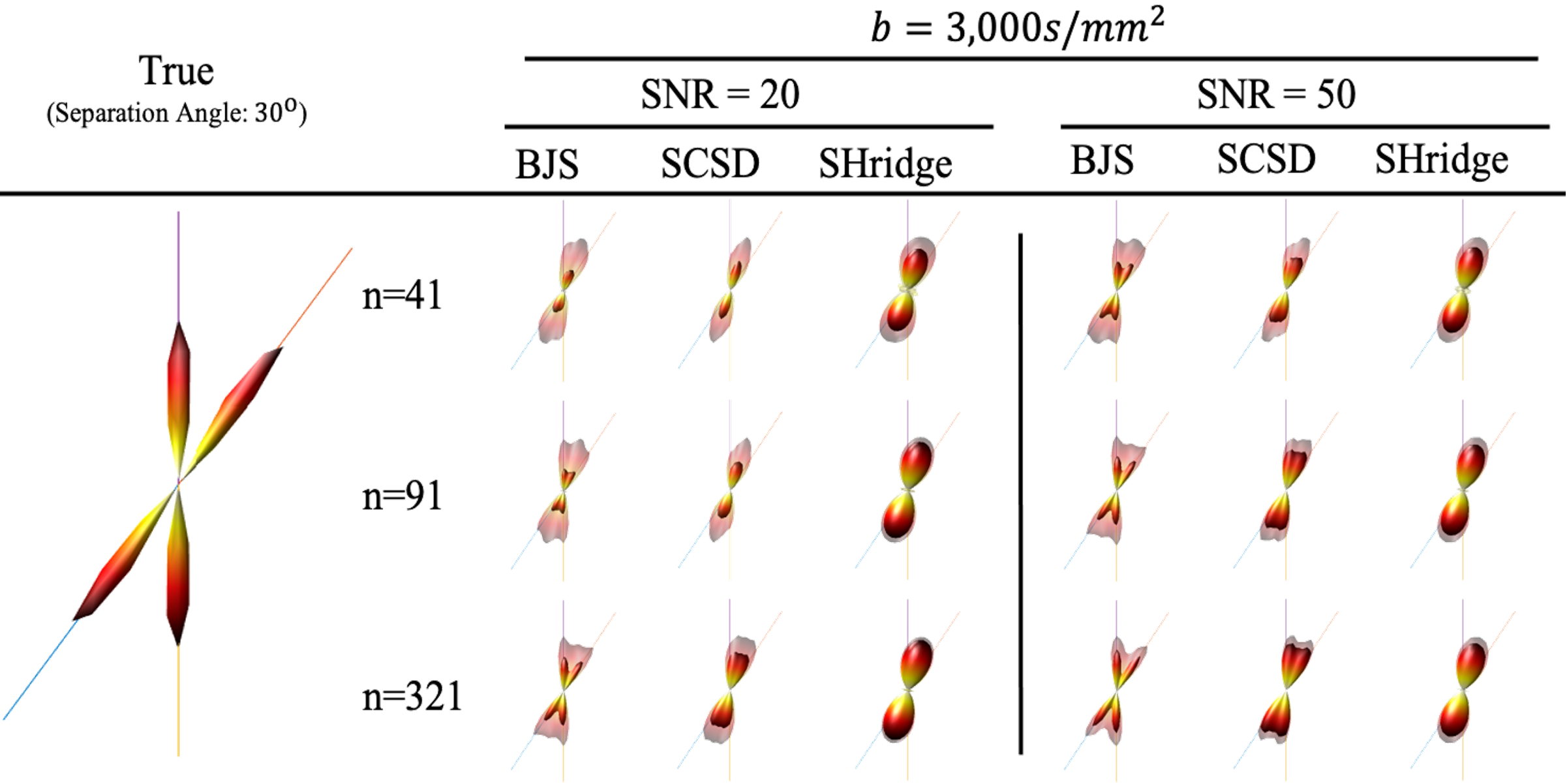}
		\caption{\textbf{Synthetic experiment: two fibers crossing at $30^{\circ}$.} The solid lines are the true fiber directions. The opaque part and the semitranslucent part represent the mean estimated FODs across 100 replicates and the mean plus two standard deviations of the estimated FODs.}
		\label{fig:simul_degree30}
	\end{figure}
	
	\begin{table}[h]
		\caption{\textbf{Synthetic experiment: two fibers crossing at  $30^{\circ}$.} \textbf{D.R.}:  correct peak detection rate;  \textbf{Bias.Sep} (\textbf{s.e.}):   bias (standard error) of separation angle estimation (in arc degree); \textbf{RMSAE}: root mean squared acute angular error (in arc degree) of fiber direction estimation.}
		\label{table:simul_degree30}
		\centering
		\resizebox{\columnwidth}{!}{%
			\begin{tabular}{ccccccccccc}
				\multicolumn{11}{l}{$b=3000s/mm^2, l_{\max}^s=16$}                                                                                                                                                                                                                                                                                                                                                                                                                                           \\ \hline
				\multicolumn{1}{|c|}{\multirow{2}{*}{\textbf{SNR}}} & \multicolumn{1}{c|}{\multirow{2}{*}{\textbf{Design}}} & \multicolumn{3}{c|}{\textbf{BJS}}                                                                                        & \multicolumn{3}{c|}{\textbf{SCSD}}                                                                                       & \multicolumn{3}{c|}{\textbf{SHridge}}                                                                                    \\ \cline{3-11} 
				\multicolumn{1}{|c|}{}                              & \multicolumn{1}{c|}{}                                 & \multicolumn{1}{c|}{\textbf{D.R.}} & \multicolumn{1}{c|}{\textbf{Bias.Sep (s.e.)}} & \multicolumn{1}{c|}{\textbf{RMSAE}} & \multicolumn{1}{c|}{\textbf{D.R.}} & \multicolumn{1}{c|}{\textbf{Bias.Sep (s.e.)}} & \multicolumn{1}{c|}{\textbf{RMSAE}} & \multicolumn{1}{c|}{\textbf{D.R.}} & \multicolumn{1}{c|}{\textbf{Bias.Sep (s.e.)}} & \multicolumn{1}{c|}{\textbf{RMSAE}} \\ \hline
				\multicolumn{1}{|c|}{\multirow{3}{*}{\textbf{20}}}  & \multicolumn{1}{c|}{$n= 41, l_{\max}=6$}              & \multicolumn{1}{c|}{46\%}          & \multicolumn{1}{c|}{1.60 (0.44)}              & \multicolumn{1}{c|}{12.03}          & \multicolumn{1}{c|}{29\%}          & \multicolumn{1}{c|}{2.47 (0.72)}              & \multicolumn{1}{c|}{11.17}          & \multicolumn{1}{c|}{3\%}           & \multicolumn{1}{c|}{33.4 (6.95)}              & \multicolumn{1}{c|}{69.9}           \\ \cline{2-11} 
				\multicolumn{1}{|c|}{}                              & \multicolumn{1}{c|}{$n= 91, l_{\max}=10$}             & \multicolumn{1}{c|}{69\%}          & \multicolumn{1}{c|}{0.40 (0.39)}              & \multicolumn{1}{c|}{9.91}           & \multicolumn{1}{c|}{43\%}          & \multicolumn{1}{c|}{-1.95 (0.62)}             & \multicolumn{1}{c|}{7.39}           & \multicolumn{1}{c|}{1\%}           & \multicolumn{1}{c|}{31.45 (-)}                & \multicolumn{1}{c|}{77.23}          \\ \cline{2-11} 
				\multicolumn{1}{|c|}{}                              & \multicolumn{1}{c|}{$n= 321, l_{\max}=12$}            & \multicolumn{1}{c|}{82\%}          & \multicolumn{1}{c|}{0.15 (0.34)}              & \multicolumn{1}{c|}{6.99}           & \multicolumn{1}{c|}{39\%}          & \multicolumn{1}{c|}{-4.26 (0.36)}             & \multicolumn{1}{c|}{5.27}           & \multicolumn{1}{c|}{0\%}           & \multicolumn{1}{c|}{-}                        & \multicolumn{1}{c|}{-}              \\ \hline
				\multicolumn{1}{|c|}{\multirow{3}{*}{\textbf{50}}}  & \multicolumn{1}{c|}{$n= 41, l_{\max}=6$}              & \multicolumn{1}{c|}{67\%}          & \multicolumn{1}{c|}{-1.02 (0.41)}             & \multicolumn{1}{c|}{6.93}           & \multicolumn{1}{c|}{28\%}          & \multicolumn{1}{c|}{-6.63 (1.00)}             & \multicolumn{1}{c|}{6.88}           & \multicolumn{1}{c|}{1\%}           & \multicolumn{1}{c|}{41.92 (2.67)}             & \multicolumn{1}{c|}{69.39}          \\ \cline{2-11} 
				\multicolumn{1}{|c|}{}                              & \multicolumn{1}{c|}{$n= 91, l_{\max}=10$}             & \multicolumn{1}{c|}{77\%}          & \multicolumn{1}{c|}{-1.18 (0.37)}             & \multicolumn{1}{c|}{5.27}           & \multicolumn{1}{c|}{47\%}          & \multicolumn{1}{c|}{-8.25 (0.65)}             & \multicolumn{1}{c|}{7.45}           & \multicolumn{1}{c|}{0\%}           & \multicolumn{1}{c|}{-}                        & \multicolumn{1}{c|}{-}              \\ \cline{2-11} 
				\multicolumn{1}{|c|}{}                              & \multicolumn{1}{c|}{$n= 321, l_{\max}=12$}            & \multicolumn{1}{c|}{88\%}          & \multicolumn{1}{c|}{-1.56 (0.38)}             & \multicolumn{1}{c|}{4.18}           & \multicolumn{1}{c|}{58\%}          & \multicolumn{1}{c|}{-9.00 (0.50)}             & \multicolumn{1}{c|}{7.68}           & \multicolumn{1}{c|}{0\%}           & \multicolumn{1}{c|}{-}                        & \multicolumn{1}{c|}{-}              \\ \hline
				\end{tabular}%
		}

	\end{table}


	
	\begin{figure}[h]
		\centering
		\includegraphics[width=0.6\textwidth]{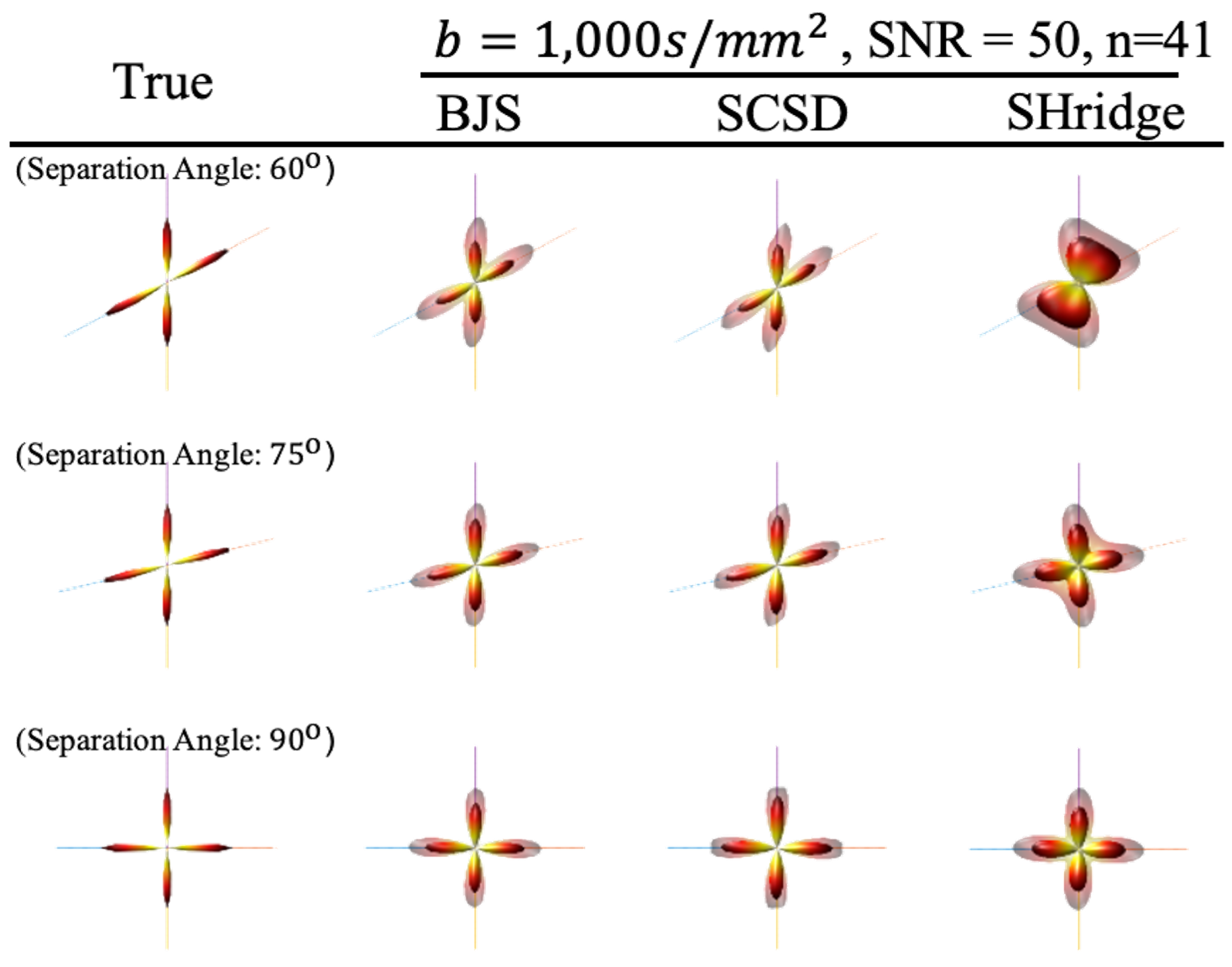}
		\caption{\textbf{Synthetic experiment: two fibers crossing at $60^{\circ}$, $75^{\circ}$, $90^{\circ}$.} The solid lines are the true fiber directions. The opaque part and the semitranslucent part represent the mean estimated FODs across 100 replicates and the mean plus two standard deviations of the estimated FODs.}
		\label{fig:simul_degree607590}
	\end{figure}
	
	\begin{table}[h]
		\caption{\textbf{Synthetic experiment:  two fibers crossing at  $60^{\circ}$, $75^{\circ}$, $90^{\circ}$.} \textbf{D.R.}:  correct peak detection rate;  \textbf{Bias.Sep} (\textbf{s.e.}):   bias (standard error) of separation angle estimation (in arc degree); \textbf{RMSAE}: root mean squared acute angular error (in arc degree) of fiber direction estimation.}
		\label{table:simul_degree607590}
		\centering
		\resizebox{\columnwidth}{!}{%
			\begin{tabular}{|c|c|c|c|c|c|c|c|c|c|}
				\multicolumn{10}{l}{$b = 1000s/mm^2, n=41, l_{\max}=6, l_{\max}^s = 12$, SNR=50}\\                                                                                                                                                                                                                                                                                                                                                                                                                                          
				\hline
				\multirow{2}{*}{\textbf{\begin{tabular}[c]{@{}c@{}}Separation\\ Angle\end{tabular}}} & \multicolumn{3}{c|}{\textbf{BJS}}                          & \multicolumn{3}{c|}{\textbf{SCSD}}                         & \multicolumn{3}{c|}{\textbf{SHridge}}                      \\ \cline{2-10} 
				                                                                                     & \textbf{D.R.} & \textbf{Bias.Sep. (s.e.)} & \textbf{RMSAE} & \textbf{D.R.} & \textbf{Bias.Sep. (s.e.)} & \textbf{RMSAE} & \textbf{D.R.} & \textbf{Bias.Sep. (s.e.)} & \textbf{RMSAE} \\ \hline
				\textbf{60}                                                                          & 93\%          & 1.24 (0.55)               & 12.66          & 89\%          & -7.68 (0.78)              & 13.06          & 42\%          & 5.90 (1.55)               & 22.00          \\ \hline
				\textbf{75}                                                                          & 98\%          & 0.09 (0.36)               & 8.94           & 83\%          & -2.49 (0.28)              & 7.85           & 77\%          & 5.17 (0.57)               & 10.94          \\ \hline
				\textbf{90}                                                                          & 100\%         & -1.83 (0.21)              & 7.62           & 88\%          & -1.53 (0.18)              & 5.96           & 93\%          & -3.24 (0.25)              & 9.00           \\ \hline
			\end{tabular}%
		}
	\end{table}


	\begin{figure}[h]
		\centering
		\includegraphics[width=0.8\textwidth]{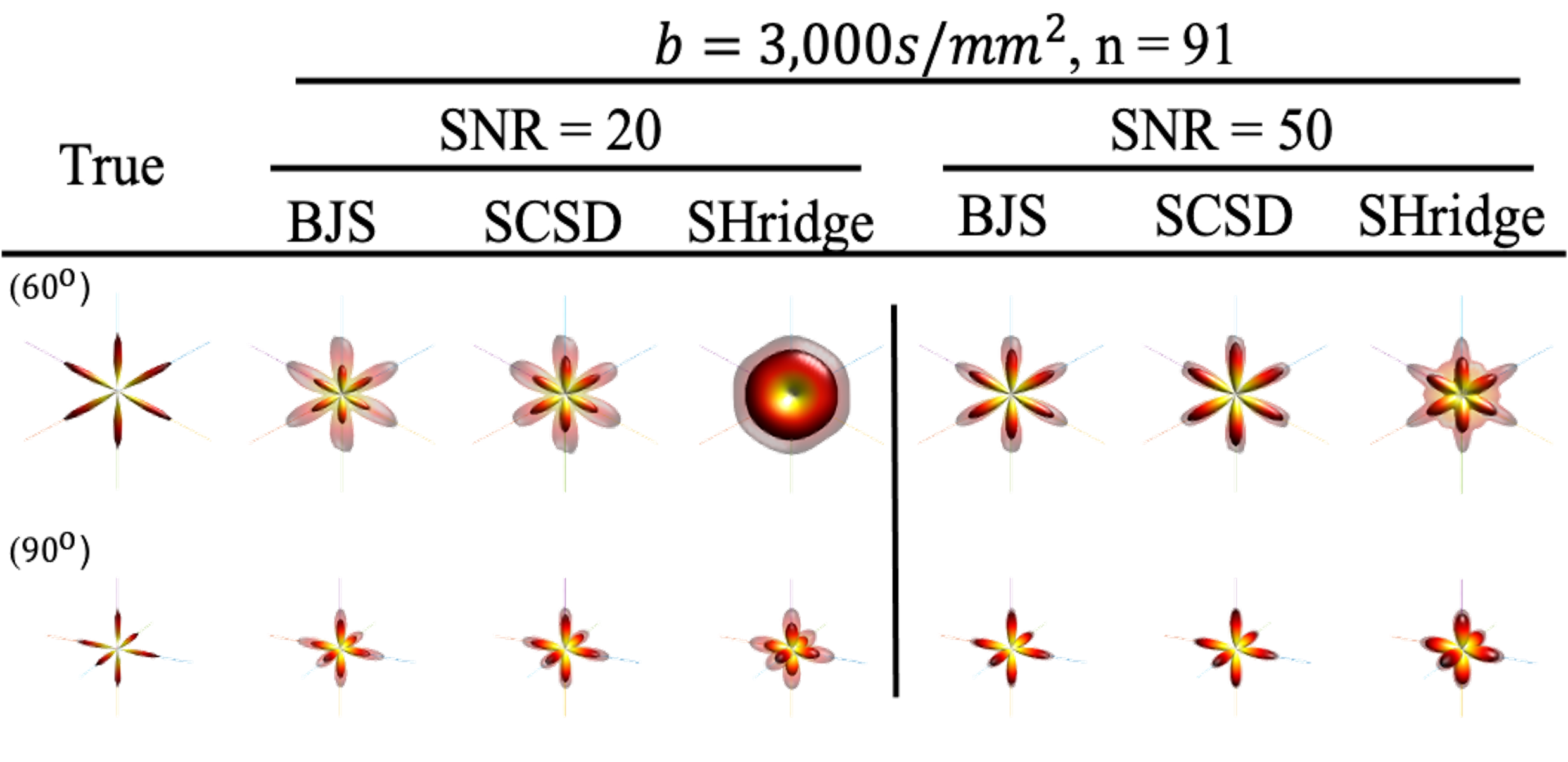}
		\caption{\textbf{Synthetic experiment: three fibers crossing at $60^{\circ}$, $90^{\circ}$.} The solid lines are the true fiber directions. The opaque part and the semitranslucent part represent the mean estimated FODs  across 100 replicates and the mean plus two standard deviations of the estimated FODs.}
		\label{fig:simul_fib3}
	\end{figure}
	
	\begin{table}[h]
		\caption{\textbf{Synthetic experiment: three fibers crossing at  ($60^{\circ}$, $90^{\circ}$).} \textbf{D.R.}:  correct peak detection rate;  \textbf{Bias.Sep}: bias of separation angle estimation (in arc degree); \textbf{RMSAE}: root mean squared acute angular error (in arc degree) of fiber direction estimation.}
		\label{table:simul_fib3}
		\centering
		\resizebox{\columnwidth}{!}{%
		\begin{tabular}{cccccccccccccccc}
			\multicolumn{16}{l}{$b=3000s/mm^2, n=91, l_{\max}=10, l_{\max}^s=12$}                                                                                                                                                                                                                                                                                                                                                                                                                                                                                                                                                                                                          \\
			\multicolumn{16}{l}{\textbf{Separation Angle: $60^{\circ}$}}                                                                                                                                                                                                                                                                                                                                                                                                                                                                                                                                                                                                                   \\ \hline
			\multicolumn{1}{|c|}{\multirow{2}{*}{\textbf{SNR}}} & \multicolumn{5}{c|}{\textbf{BJS}}                                                                                                                                                                      & \multicolumn{5}{c|}{\textbf{SCSD}}                                                                                                                                                                     & \multicolumn{5}{c|}{\textbf{SHridge}}                                                                                                                                                                  \\ \cline{2-16} 
			\multicolumn{1}{|c|}{}                              & \multicolumn{1}{c|}{\textbf{D.R.}} & \multicolumn{1}{c|}{\textbf{Bias.Sep1}} & \multicolumn{1}{c|}{\textbf{Bias.Sep2}} & \multicolumn{1}{c|}{\textbf{Bias.Sep3}} & \multicolumn{1}{c|}{\textbf{RMSAE}} & \multicolumn{1}{c|}{\textbf{D.R.}} & \multicolumn{1}{c|}{\textbf{Bias.Sep1}} & \multicolumn{1}{c|}{\textbf{Bias.Sep2}} & \multicolumn{1}{c|}{\textbf{Bias.Sep3}} & \multicolumn{1}{c|}{\textbf{RMSAE}} & \multicolumn{1}{c|}{\textbf{D.R.}} & \multicolumn{1}{c|}{\textbf{Bias.Sep1}} & \multicolumn{1}{c|}{\textbf{Bias.Sep2}} & \multicolumn{1}{c|}{\textbf{Bias.Sep3}} & \multicolumn{1}{c|}{\textbf{RMSAE}} \\ \hline
			\multicolumn{1}{|c|}{\textbf{20}}                   & \multicolumn{1}{c|}{71\%}          & \multicolumn{1}{c|}{-0.39}              & \multicolumn{1}{c|}{0.61}               & \multicolumn{1}{c|}{-0.75}              & \multicolumn{1}{c|}{14.4}           & \multicolumn{1}{c|}{38\%}          & \multicolumn{1}{c|}{-0.64}              & \multicolumn{1}{c|}{0.39}               & \multicolumn{1}{c|}{0.64}               & \multicolumn{1}{c|}{14.03}          & \multicolumn{1}{c|}{0\%}           & \multicolumn{1}{c|}{-}                  & \multicolumn{1}{c|}{-}                  & \multicolumn{1}{c|}{-}                  & \multicolumn{1}{c|}{-}              \\ \hline
			\multicolumn{1}{|c|}{\textbf{50}}                   & \multicolumn{1}{c|}{93\%}          & \multicolumn{1}{c|}{-1.22}              & \multicolumn{1}{c|}{0.48}               & \multicolumn{1}{c|}{0.81}               & \multicolumn{1}{c|}{7.28}           & \multicolumn{1}{c|}{99\%}          & \multicolumn{1}{c|}{0.11}               & \multicolumn{1}{c|}{0.27}               & \multicolumn{1}{c|}{-0.33}              & \multicolumn{1}{c|}{5.66}           & \multicolumn{1}{c|}{67\%}          & \multicolumn{1}{c|}{-0.49}              & \multicolumn{1}{c|}{1.08}               & \multicolumn{1}{c|}{0.63}               & \multicolumn{1}{c|}{17.23}          \\ \hline
			\multicolumn{16}{l}{}                                                                                                                                                                                                                                                                                                                                                                                                                                                                                                                                                                                                                                                          \\
			\multicolumn{16}{l}{\textbf{Separation Angle: $90^{\circ}$}}                                                                                                                                                                                                                                                                                                                                                                                                                                                                                                                                                                                                                   \\ \hline
			\multicolumn{1}{|c|}{\multirow{2}{*}{\textbf{SNR}}} & \multicolumn{5}{c|}{\textbf{BJS}}                                                                                                                                                                      & \multicolumn{5}{c|}{\textbf{SCSD}}                                                                                                                                                                     & \multicolumn{5}{c|}{\textbf{SHridge}}                                                                                                                                                                  \\ \cline{2-16} 
			\multicolumn{1}{|c|}{}                              & \multicolumn{1}{c|}{\textbf{D.R.}} & \multicolumn{1}{c|}{\textbf{Bias.Sep1}} & \multicolumn{1}{c|}{\textbf{Bias.Sep2}} & \multicolumn{1}{c|}{\textbf{Bias.Sep3}} & \multicolumn{1}{c|}{\textbf{RMSAE}} & \multicolumn{1}{c|}{\textbf{D.R.}} & \multicolumn{1}{c|}{\textbf{Bias.Sep1}} & \multicolumn{1}{c|}{\textbf{Bias.Sep2}} & \multicolumn{1}{c|}{\textbf{Bias.Sep3}} & \multicolumn{1}{c|}{\textbf{RMSAE}} & \multicolumn{1}{c|}{\textbf{D.R.}} & \multicolumn{1}{c|}{\textbf{Bias.Sep1}} & \multicolumn{1}{c|}{\textbf{Bias.Sep2}} & \multicolumn{1}{c|}{\textbf{Bias.Sep3}} & \multicolumn{1}{c|}{\textbf{RMSAE}} \\ \hline
			\multicolumn{1}{|c|}{\textbf{20}}                   & \multicolumn{1}{c|}{88\%}          & \multicolumn{1}{c|}{-1.68}              & \multicolumn{1}{c|}{-2.93}              & \multicolumn{1}{c|}{-2.25}              & \multicolumn{1}{c|}{7.94}           & \multicolumn{1}{c|}{100\%}         & \multicolumn{1}{c|}{-1.74}              & \multicolumn{1}{c|}{-2.39}              & \multicolumn{1}{c|}{-2.26}              & \multicolumn{1}{c|}{7.16}           & \multicolumn{1}{c|}{90\%}          & \multicolumn{1}{c|}{-2.04}              & \multicolumn{1}{c|}{-2.30}              & \multicolumn{1}{c|}{-2.62}              & \multicolumn{1}{c|}{7.22}           \\ \hline
			\multicolumn{1}{|c|}{\textbf{50}}                   & \multicolumn{1}{c|}{94\%}          & \multicolumn{1}{c|}{-0.20}              & \multicolumn{1}{c|}{-0.34}              & \multicolumn{1}{c|}{-0.45}              & \multicolumn{1}{c|}{2.26}           & \multicolumn{1}{c|}{100\%}         & \multicolumn{1}{c|}{-0.10}              & \multicolumn{1}{c|}{-0.35}              & \multicolumn{1}{c|}{-0.23}              & \multicolumn{1}{c|}{1.49}           & \multicolumn{1}{c|}{100\%}         & \multicolumn{1}{c|}{-0.59}              & \multicolumn{1}{c|}{-1.62}              & \multicolumn{1}{c|}{-2.03}              & \multicolumn{1}{c|}{2.80}           \\ \hline
			\end{tabular}%
		}

	\end{table}
	
	\begin{table}[h]
		\caption{\textbf{Sensitivity experiment: two fibers crossing at  $45^{\circ}$ with 5 different choices of the shrinkage parameter $t^{(l)}=c\log(2l+1)$  in \textbf{BJS}.} \textbf{D.R.}:  correct peak detection rate;  \textbf{Bias.Sep} (\textbf{s.e.}):   bias (standard error) of separation angle estimation (in arc degree); \textbf{RMSAE}: root mean squared acute angular error (in arc degree) of fiber direction estimation. }
		\label{table:sensitivity_degree45}
		\centering
		\resizebox{\columnwidth}{!}{%
			\begin{tabular}{|c|c|c|c|c|c|c|c|c|c|c|}
				\hline
				\multirow{3}{*}{\textbf{Setting}}                                                & \multirow{3}{*}{\textbf{c}} & \multicolumn{3}{c|}{$b=1000s/mm^2$, $l_{\max}^s = 12$}                        & \multicolumn{6}{c|}{$b=3000s/mm^2$, $l_{\max}^s = 12$}                                                                                     \\ \cline{3-11} 
				                                                                                 &                             & \multicolumn{3}{c|}{\textbf{SNR=50}}                       & \multicolumn{3}{c|}{\textbf{SNR=20}}                       & \multicolumn{3}{c|}{\textbf{SNR=50}}                       \\ \cline{3-11} 
				                                                                                 &                             & \textbf{D.R.} & \textbf{Bias.Sep. (s.e.)} & \textbf{RMSAE} & \textbf{D.R.} & \textbf{Bias.Sep. (s.e.)} & \textbf{RMSAE} & \textbf{D.R.} & \textbf{Bias.Sep. (s.e.)} & \textbf{RMSAE} \\ \hline
				\multirow{5}{*}{\begin{tabular}[c]{@{}c@{}}$n=41$\\ $l_{\max}=6$\end{tabular}}   & \textbf{1}                  & 62\%          & -1.38 (0.78)              & 10.89          & 91\%          & -1.68 (0.62)              & 11.34          & 100\%         & -1.12 (0.29)              & 5.16           \\ \cline{2-11} 
				                                                                                 & \textbf{1.5}                & 62\%          & -1.33 (0.78)              & 10.89          & 91\%          & -1.72 (0.62)              & 11.34          & 100\%         & -1.08 (0.29)              & 5.10           \\ \cline{2-11} 
				                                                                                 & \textbf{2}                  & 62\%          & -1.33 (0.78)              & 10.89          & 92\%          & -1.71 (0.61)              & 11.40          & 100\%         & -1.03 (0.29)              & 5.10           \\ \cline{2-11} 
				                                                                                 & \textbf{2.5}                & 62\%          & -1.33 (0.78)              & 10.89          & 92\%          & -1.73 (0.61)              & 11.34          & 100\%         & -1 (0.29)                 & 5.10           \\ \cline{2-11} 
				                                                                                 & \textbf{3}                  & 62\%          & -1.33 (0.78)              & 10.89          & 92\%          & -1.77 (0.61)              & 11.34          & 100\%         & -0.96 (0.29)              & 5.10           \\ \hline
				\multirow{5}{*}{\begin{tabular}[c]{@{}c@{}}$n=91$\\ $l_{\max}=10$\end{tabular}}  & \textbf{1}                  & 77\%          & -0.56 (0.63)              & 10.54          & 94\%          & -1.79 (0.33)              & 8.42           & 96\%          & -0.53 (0.23)              & 3.78           \\ \cline{2-11} 
				                                                                                 & \textbf{1.5}                & 82\%          & -0.74 (0.63)              & 10.49          & 97\%          & -1.64 (0.31)              & 7.79           & 98\%          & -0.2 (0.24)               & 3.32           \\ \cline{2-11} 
				                                                                                 & \textbf{2}                  & 83\%          & -0.59 (0.59)              & 10.37          & 97\%          & -1.67 (0.31)              & 7.79           & 98\%          & -0.05 (0.2)               & 3.21           \\ \cline{2-11} 
				                                                                                 & \textbf{2.5}                & 84\%          & -0.56 (0.58)              & 10.37          & 98\%          & -1.62 (0.31)              & 7.62           & 99\%          & -0.11 (0.19)              & 3.15           \\ \cline{2-11} 
				                                                                                 & \textbf{3}                  & 85\%          & -0.63 (0.58)              & 10.37          & 98\%          & -1.62 (0.31)              & 7.62           & 100\%         & -0.28 (0.2)               & 3.27           \\ \hline
				\multirow{5}{*}{\begin{tabular}[c]{@{}c@{}}$n=321$\\ $l_{\max}=12$\end{tabular}} & \textbf{1}                  & 96\%          & -0.17 (0.32)              & 6.13           & 98\%          & -1.98 (0.24)              & 5.44           & 97\%          & -0.02 (0.16)              & 2.01           \\ \cline{2-11} 
				                                                                                 & \textbf{1.5}                & 99\%          & -0.16 (0.31)              & 6.07           & 99\%          & -1.9 (0.23)               & 5.73           & 99\%          & -0.14 (0.17)              & 2.12           \\ \cline{2-11} 
				                                                                                 & \textbf{2}                  & 100\%         & -0.16 (0.31)              & 6.13           & 100\%         & -1.99 (0.22)              & 5.73           & 100\%         & -0.34 (0.17)              & 2.29           \\ \cline{2-11} 
				                                                                                 & \textbf{2.5}                & 100\%         & -0.16 (0.31)              & 6.13           & 100\%         & -2.11 (0.23)              & 5.73           & 100\%         & -0.49 (0.18)              & 2.41           \\ \cline{2-11} 
				                                                                                 & \textbf{3}                  & 100\%         & -0.16 (0.31)              & 6.13           & 100\%         & -2.03 (0.23)              & 5.73           & 100\%         & -0.46 (0.18)              & 2.41           \\ \hline
			\end{tabular}%
		}
	\end{table}

	\begin{table}[h]
		\caption{\textbf{Sensitivity experiment: two fibers crossing  at $45^{\circ}$ with 4 different $l^s_{\max}$ in the post-estimation sharpening step of \textbf{BJS}.} \textbf{D.R.}:  correct peak detection rate;  \textbf{Bias.Sep} (\textbf{s.e.}):   bias (standard error) of separation angle estimation (in arc degree); \textbf{RMSAE}: root mean squared acute angular error (in arc degree) of fiber direction estimation.}
		\label{table:sensitivity_lmaxs}
		\centering
		\resizebox{\columnwidth}{!}{%
			\begin{tabular}{cccccccccccccc}
				\multicolumn{14}{l}{$b=1000s/mm^2$}                                                                                                                                                                                                                                                                                                                                                                                                                                                                                                                                                                                         \\ \hline
				\multicolumn{1}{|c|}{\multirow{2}{*}{\textbf{SNR}}} & \multicolumn{1}{c|}{\multirow{2}{*}{\textbf{Design}}} & \multicolumn{3}{c|}{$l_{\max}^s=10$}                                                                                      & \multicolumn{3}{c|}{$l_{\max}^s=12$}                                                                                      & \multicolumn{3}{c|}{$l_{\max}^s=14$}                                                                                      & \multicolumn{3}{c|}{\textbf{$l_{\max}^s=16$}}                                                                             \\ \cline{3-14} 
				\multicolumn{1}{|c|}{}                              & \multicolumn{1}{c|}{}                                 & \multicolumn{1}{c|}{\textbf{D.R.}} & \multicolumn{1}{c|}{\textbf{Bias.Sep. (s.e.)}} & \multicolumn{1}{c|}{\textbf{RMSAE}} & \multicolumn{1}{c|}{\textbf{D.R.}} & \multicolumn{1}{c|}{\textbf{Bias.Sep. (s.e.)}} & \multicolumn{1}{c|}{\textbf{RMSAE}} & \multicolumn{1}{c|}{\textbf{D.R.}} & \multicolumn{1}{c|}{\textbf{Bias.Sep. (s.e.)}} & \multicolumn{1}{c|}{\textbf{RMSAE}} & \multicolumn{1}{c|}{\textbf{D.R.}} & \multicolumn{1}{c|}{\textbf{Bias.Sep. (s.e.)}} & \multicolumn{1}{c|}{\textbf{RMSAE}} \\ \hline
				\multicolumn{1}{|c|}{\multirow{3}{*}{\textbf{50}}}  & \multicolumn{1}{c|}{$n= 41, l_{\max}=6$}              & \multicolumn{1}{c|}{18\%}          & \multicolumn{1}{c|}{-5.87 (1.33)}              & \multicolumn{1}{c|}{75.92}          & \multicolumn{1}{c|}{62\%}          & \multicolumn{1}{c|}{-1.33 (0.78)}              & \multicolumn{1}{c|}{44.63}          & \multicolumn{1}{c|}{69\%}          & \multicolumn{1}{c|}{-0.57 (0.69)}              & \multicolumn{1}{c|}{39.48}          & \multicolumn{1}{c|}{67\%}          & \multicolumn{1}{c|}{-3.88 (0.76)}              & \multicolumn{1}{c|}{43.6}           \\ \cline{2-14} 
				\multicolumn{1}{|c|}{}                              & \multicolumn{1}{c|}{$n= 91, l_{\max}=10$}             & \multicolumn{1}{c|}{29\%}          & \multicolumn{1}{c|}{-3.58 (0.94)}              & \multicolumn{1}{c|}{54.09}          & \multicolumn{1}{c|}{83\%}          & \multicolumn{1}{c|}{-0.59 (0.59)}              & \multicolumn{1}{c|}{33.75}          & \multicolumn{1}{c|}{88\%}          & \multicolumn{1}{c|}{-2.05 (0.56)}              & \multicolumn{1}{c|}{31.91}          & \multicolumn{1}{c|}{88\%}          & \multicolumn{1}{c|}{-6.15 (0.45)}              & \multicolumn{1}{c|}{25.73}          \\ \cline{2-14} 
				\multicolumn{1}{|c|}{}                              & \multicolumn{1}{c|}{$n= 321, l_{\max}=12$}            & \multicolumn{1}{c|}{-}             & \multicolumn{1}{c|}{-}                         & \multicolumn{1}{c|}{-}              & \multicolumn{1}{c|}{100\%}         & \multicolumn{1}{c|}{-0.16 (0.31)}              & \multicolumn{1}{c|}{17.76}          & \multicolumn{1}{c|}{100\%}         & \multicolumn{1}{c|}{-2.77 (0.38)}              & \multicolumn{1}{c|}{21.49}          & \multicolumn{1}{c|}{89\%}          & \multicolumn{1}{c|}{-6.23 (0.34)}              & \multicolumn{1}{c|}{19.71}          \\ \hline
				&                                                       &                                    &                                                &                                     &                                    &                                                &                                     &                                    &                                                &                                     &                                    &                                                &                                     \\
				\multicolumn{14}{l}{$b=3000s/mm^2$}                                                                                                                                                                                                                                                                                                                                                                                                                                                                                                                                                                                         \\ \hline
				\multicolumn{1}{|c|}{\multirow{2}{*}{\textbf{SNR}}} & \multicolumn{1}{c|}{\multirow{2}{*}{\textbf{Design}}} & \multicolumn{3}{c|}{$l_{\max}^s=10$}                                                                                      & \multicolumn{3}{c|}{$l_{\max}^s=12$}                                                                                      & \multicolumn{3}{c|}{$l_{\max}^s=14$}                                                                                      & \multicolumn{3}{c|}{\textbf{$l_{\max}^s=16$}}                                                                             \\ \cline{3-14} 
				\multicolumn{1}{|c|}{}                              & \multicolumn{1}{c|}{}                                 & \multicolumn{1}{c|}{\textbf{D.R.}} & \multicolumn{1}{c|}{\textbf{Bias.Sep. (s.e.)}} & \multicolumn{1}{c|}{\textbf{RMSAE}} & \multicolumn{1}{c|}{\textbf{D.R.}} & \multicolumn{1}{c|}{\textbf{Bias.Sep. (s.e.)}} & \multicolumn{1}{c|}{\textbf{RMSAE}} & \multicolumn{1}{c|}{\textbf{D.R.}} & \multicolumn{1}{c|}{\textbf{Bias.Sep. (s.e.)}} & \multicolumn{1}{c|}{\textbf{RMSAE}} & \multicolumn{1}{c|}{\textbf{D.R.}} & \multicolumn{1}{c|}{\textbf{Bias.Sep. (s.e.)}} & \multicolumn{1}{c|}{\textbf{RMSAE}} \\ \hline
				\multicolumn{1}{|c|}{\multirow{3}{*}{\textbf{20}}}  & \multicolumn{1}{c|}{$n= 41, l_{\max}=6$}              & \multicolumn{1}{c|}{67\%}          & \multicolumn{1}{c|}{-2.22 (0.78)}              & \multicolumn{1}{c|}{44.92}          & \multicolumn{1}{c|}{92\%}          & \multicolumn{1}{c|}{-1.71 (0.61)}              & \multicolumn{1}{c|}{35.07}          & \multicolumn{1}{c|}{91\%}          & \multicolumn{1}{c|}{-4.16 (0.54)}              & \multicolumn{1}{c|}{31.05}          & \multicolumn{1}{c|}{71\%}          & \multicolumn{1}{c|}{-8.32 (0.48)}              & \multicolumn{1}{c|}{27.44}          \\ \cline{2-14} 
				\multicolumn{1}{|c|}{}                              & \multicolumn{1}{c|}{$n= 91, l_{\max}=10$}             & \multicolumn{1}{c|}{90\%}          & \multicolumn{1}{c|}{-1.2 (0.44)}               & \multicolumn{1}{c|}{25.44}          & \multicolumn{1}{c|}{97\%}          & \multicolumn{1}{c|}{-1.67 (0.31)}              & \multicolumn{1}{c|}{17.82}          & \multicolumn{1}{c|}{88\%}          & \multicolumn{1}{c|}{-4.38 (0.36)}              & \multicolumn{1}{c|}{20.63}          & \multicolumn{1}{c|}{68\%}          & \multicolumn{1}{c|}{-7.09 (0.54)}              & \multicolumn{1}{c|}{30.83}          \\ \cline{2-14} 
				\multicolumn{1}{|c|}{}                              & \multicolumn{1}{c|}{$n= 321, l_{\max}=12$}            & \multicolumn{1}{c|}{-}             & \multicolumn{1}{c|}{-}                         & \multicolumn{1}{c|}{-}              & \multicolumn{1}{c|}{100\%}         & \multicolumn{1}{c|}{-1.99 (0.22)}              & \multicolumn{1}{c|}{12.72}          & \multicolumn{1}{c|}{96\%}          & \multicolumn{1}{c|}{-4.34 (0.36)}              & \multicolumn{1}{c|}{20.74}          & \multicolumn{1}{c|}{67\%}          & \multicolumn{1}{c|}{-8.08 (0.44)}              & \multicolumn{1}{c|}{25.04}          \\ \hline
				\multicolumn{1}{|c|}{\multirow{3}{*}{\textbf{50}}}  & \multicolumn{1}{c|}{$n= 41, l_{\max}=6$}              & \multicolumn{1}{c|}{96\%}          & \multicolumn{1}{c|}{-3.25 (0.46)}              & \multicolumn{1}{c|}{26.47}          & \multicolumn{1}{c|}{100\%}         & \multicolumn{1}{c|}{-1.03 (0.29)}              & \multicolumn{1}{c|}{16.56}          & \multicolumn{1}{c|}{100\%}         & \multicolumn{1}{c|}{-3.58 (0.30)}              & \multicolumn{1}{c|}{16.96}          & \multicolumn{1}{c|}{83\%}          & \multicolumn{1}{c|}{-6.04 (0.47)}              & \multicolumn{1}{c|}{26.64}          \\ \cline{2-14} 
				\multicolumn{1}{|c|}{}                              & \multicolumn{1}{c|}{$n= 91, l_{\max}=10$}             & \multicolumn{1}{c|}{98\%}          & \multicolumn{1}{c|}{-0.69 (0.25)}              & \multicolumn{1}{c|}{14.21}          & \multicolumn{1}{c|}{98\%}          & \multicolumn{1}{c|}{-0.05 (0.20)}              & \multicolumn{1}{c|}{11.46}          & \multicolumn{1}{c|}{97\%}          & \multicolumn{1}{c|}{-2.71 (0.28)}              & \multicolumn{1}{c|}{16.21}          & \multicolumn{1}{c|}{83\%}          & \multicolumn{1}{c|}{-4.79 (0.50)}              & \multicolumn{1}{c|}{28.42}          \\ \cline{2-14} 
				\multicolumn{1}{|c|}{}                              & \multicolumn{1}{c|}{$n= 321, l_{\max}=12$}            & \multicolumn{1}{c|}{-}             & \multicolumn{1}{c|}{-}                         & \multicolumn{1}{c|}{-}              & \multicolumn{1}{c|}{100\%}         & \multicolumn{1}{c|}{-0.34 (0.17)}              & \multicolumn{1}{c|}{9.85}           & \multicolumn{1}{c|}{100\%}         & \multicolumn{1}{c|}{-0.85 (0.20)}              & \multicolumn{1}{c|}{11.69}          & \multicolumn{1}{c|}{100\%}         & \multicolumn{1}{c|}{-2.05 (0.26)}              & \multicolumn{1}{c|}{14.84}          \\ \hline
			\end{tabular}%
		}
	\end{table}

	\begin{table}[h]
		\caption{\textbf{Sensitivity experiment:  two fibers crossing at $45^{\circ}$ with three different levels of suppressing small positive values in the post-estimation sharpening step of \textbf{BJS}.}  $thre_p$ means to suppress values less than $thre_p$ times the mean of the initial FOD estimate. 
			\textbf{D.R.}:  correct peak detection rate;  \textbf{Bias.Sep} (\textbf{s.e.}):   bias (standard error) of separation angle estimation (in arc degree); \textbf{RMSAE}: root mean squared acute angular error (in arc degree) of fiber direction estimation. }
			\label{table:sharpening}
		\centering
		\resizebox{\columnwidth}{!}{%
			\begin{tabular}{ccccccccccc}
				\multicolumn{11}{l}{$b=1000s/mm^2, l_{\max}^s=12$}                                                                                                                                                                                                                                                                                                                                                                                                                                              \\ \hline
				\multicolumn{1}{|c|}{\multirow{2}{*}{\textbf{SNR}}} & \multicolumn{1}{c|}{\multirow{2}{*}{\textbf{Design}}} & \multicolumn{3}{c|}{$thre_p=0$}                                                                                         & \multicolumn{3}{c|}{$thre_p=0.1$}                                                                                   & \multicolumn{3}{c|}{$thre_p=0.5$}                                                                                   \\ \cline{3-11} 
				\multicolumn{1}{|c|}{}                              & \multicolumn{1}{c|}{}                                 & \multicolumn{1}{c|}{\textbf{D.R.}} & \multicolumn{1}{c|}{\textbf{Bias.Sep. (s.e.)}} & \multicolumn{1}{c|}{\textbf{RMSAE}} & \multicolumn{1}{c|}{\textbf{D.R.}} & \multicolumn{1}{c|}{\textbf{Bias.Sep. (s.e.)}} & \multicolumn{1}{c|}{\textbf{RMSAE}} & \multicolumn{1}{c|}{\textbf{D.R.}} & \multicolumn{1}{c|}{\textbf{Bias.Sep. (s.e.)}} & \multicolumn{1}{c|}{\textbf{RMSAE}} \\ \hline
				\multicolumn{1}{|c|}{\multirow{3}{*}{\textbf{50}}}  & \multicolumn{1}{c|}{$n= 41, l_{\max}=6$}              & \multicolumn{1}{c|}{62\%}          & \multicolumn{1}{c|}{-1.33 (0.78)}              & \multicolumn{1}{c|}{10.89}          & \multicolumn{1}{c|}{61\%}          & \multicolumn{1}{c|}{-1.45 (0.75)}              & \multicolumn{1}{c|}{10.26}          & \multicolumn{1}{c|}{54\%}          & \multicolumn{1}{c|}{-4.32 (0.78)}              & \multicolumn{1}{c|}{9.97}           \\ \cline{2-11} 
				\multicolumn{1}{|c|}{}                              & \multicolumn{1}{c|}{$n= 91, l_{\max}=10$}             & \multicolumn{1}{c|}{83\%}          & \multicolumn{1}{c|}{-0.59 (0.59)}              & \multicolumn{1}{c|}{10.37}          & \multicolumn{1}{c|}{83\%}          & \multicolumn{1}{c|}{-1.15 (0.61)}              & \multicolumn{1}{c|}{9.57}           & \multicolumn{1}{c|}{83\%}          & \multicolumn{1}{c|}{-3.33 (0.58)}              & \multicolumn{1}{c|}{8.14}           \\ \cline{2-11} 
				\multicolumn{1}{|c|}{}                              & \multicolumn{1}{c|}{$n= 321, l_{\max}=12$}            & \multicolumn{1}{c|}{100\%}         & \multicolumn{1}{c|}{-0.16 (0.31)}              & \multicolumn{1}{c|}{6.13}           & \multicolumn{1}{c|}{100\%}         & \multicolumn{1}{c|}{-0.14 (0.28)}              & \multicolumn{1}{c|}{5.79}           & \multicolumn{1}{c|}{100\%}         & \multicolumn{1}{c|}{-1.97 (0.30)}              & \multicolumn{1}{c|}{5.16}           \\ \hline
				\multicolumn{11}{l}{}                                                                                                                                                                                                                                                                                                                                                                                                                                                                           \\
				\multicolumn{11}{l}{$b=3000s/mm^2, l_{\max}^s=12$}                                                                                                                                                                                                                                                                                                                                                                                                                                              \\ \hline
				\multicolumn{1}{|c|}{\multirow{2}{*}{\textbf{SNR}}} & \multicolumn{1}{c|}{\multirow{2}{*}{\textbf{Design}}} & \multicolumn{3}{c|}{$thre_p=0$}                                                                                         & \multicolumn{3}{c|}{$thre_p=0.1$}                                                                                   & \multicolumn{3}{c|}{$thre_p=0.5$}                                                                                                                                                                          \\ \cline{3-11} 
				\multicolumn{1}{|c|}{}                              & \multicolumn{1}{c|}{}                                 & \multicolumn{1}{c|}{\textbf{D.R.}} & \multicolumn{1}{c|}{\textbf{Bias.Sep. (s.e.)}} & \multicolumn{1}{c|}{\textbf{RMSAE}} & \multicolumn{1}{c|}{\textbf{D.R.}} & \multicolumn{1}{c|}{\textbf{Bias.Sep. (s.e.)}} & \multicolumn{1}{c|}{\textbf{RMSAE}} & \multicolumn{1}{c|}{\textbf{D.R.}} & \multicolumn{1}{c|}{\textbf{Bias.Sep. (s.e.)}} & \multicolumn{1}{c|}{\textbf{RMSAE}} \\ \hline
				\multicolumn{1}{|c|}{\multirow{3}{*}{\textbf{20}}}  & \multicolumn{1}{c|}{$n= 41, l_{\max}=6$}              & \multicolumn{1}{c|}{92\%}          & \multicolumn{1}{c|}{-1.71 (0.61)}              & \multicolumn{1}{c|}{11.4}           & \multicolumn{1}{c|}{92\%}          & \multicolumn{1}{c|}{-1.54 (0.62)}              & \multicolumn{1}{c|}{11.34}          & \multicolumn{1}{c|}{85\%}          & \multicolumn{1}{c|}{-2.24 (0.52)}              & \multicolumn{1}{c|}{9.05}           \\ \cline{2-11} 
				\multicolumn{1}{|c|}{}                              & \multicolumn{1}{c|}{$n= 91, l_{\max}=10$}             & \multicolumn{1}{c|}{97\%}          & \multicolumn{1}{c|}{-1.67 (0.31)}              & \multicolumn{1}{c|}{7.79}           & \multicolumn{1}{c|}{97\%}          & \multicolumn{1}{c|}{-1.37 (0.34)}              & \multicolumn{1}{c|}{7.62}           & \multicolumn{1}{c|}{94\%}          & \multicolumn{1}{c|}{-1.79 (0.34)}              & \multicolumn{1}{c|}{6.53}           \\ \cline{2-11} 
				\multicolumn{1}{|c|}{}                              & \multicolumn{1}{c|}{$n= 321, l_{\max}=12$}            & \multicolumn{1}{c|}{100\%}         & \multicolumn{1}{c|}{-1.99 (0.22)}              & \multicolumn{1}{c|}{5.73}           & \multicolumn{1}{c|}{100\%}         & \multicolumn{1}{c|}{-2.29 (0.23)}              & \multicolumn{1}{c|}{5.44}           & \multicolumn{1}{c|}{100\%}         & \multicolumn{1}{c|}{-1.94 (0.21)}              & \multicolumn{1}{c|}{4.41}           \\ \hline
				\multicolumn{1}{|c|}{\multirow{3}{*}{\textbf{50}}}  & \multicolumn{1}{c|}{$n= 41, l_{\max}=6$}              & \multicolumn{1}{c|}{100\%}         & \multicolumn{1}{c|}{-1.03 (0.29)}              & \multicolumn{1}{c|}{5.10}           & \multicolumn{1}{c|}{100\%}         & \multicolumn{1}{c|}{-1.14 (0.29)}              & \multicolumn{1}{c|}{4.76}           & \multicolumn{1}{c|}{100\%}         & \multicolumn{1}{c|}{-1.99 (0.29)}              & \multicolumn{1}{c|}{4.35}           \\ \cline{2-11} 
				\multicolumn{1}{|c|}{}                              & \multicolumn{1}{c|}{$n= 91, l_{\max}=10$}             & \multicolumn{1}{c|}{98\%}          & \multicolumn{1}{c|}{-0.05 (0.20)}              & \multicolumn{1}{c|}{3.21}           & \multicolumn{1}{c|}{98\%}          & \multicolumn{1}{c|}{-0.31 (0.20)}              & \multicolumn{1}{c|}{3.15}           & \multicolumn{1}{c|}{98\%}          & \multicolumn{1}{c|}{-0.81 (0.20)}              & \multicolumn{1}{c|}{3.15}           \\ \cline{2-11} 
				\multicolumn{1}{|c|}{}                              & \multicolumn{1}{c|}{$n= 321, l_{\max}=12$}            & \multicolumn{1}{c|}{100\%}         & \multicolumn{1}{c|}{-0.34 (0.17)}              & \multicolumn{1}{c|}{2.29}           & \multicolumn{1}{c|}{100\%}         & \multicolumn{1}{c|}{-0.38 (0.17)}              & \multicolumn{1}{c|}{2.29}           & \multicolumn{1}{c|}{100\%}         & \multicolumn{1}{c|}{-0.34 (0.17)}              & \multicolumn{1}{c|}{2.23}           \\ \hline
			\end{tabular}%
		}
	\end{table}

	\clearpage
	\newpage
	
	\section{HCP Application: Additional Details and Plots}
	\subsection{Response function estimation}
	\label{sec:supp-response}
	\begin{itemize}
		\item[\textbf{Step 1}] At each white-matter voxel, first fit the single tensor model (\ref{eq:single-tensor}); then calculate the FA value (\ref{eq:FA}) and  the ratio between the two smaller eigenvalues of the diffusion tensor $\mathbf{D}$. 
		\item[\textbf{Step 2}] Identify voxels with a single dominant fiber bundle characterized by $FA>0.8$ and the ratio between the two smaller eigenvalues $<1.5$. Define the minor eigenvalue as the average of the two smaller eigenvalues.
		\item[\textbf{Step 3}] Calculate the median of the leading eigenvalue and the minor eigenvalue across voxels from Step 2, denoted by $\bar{\lambda}$ and $\underline{\lambda}$, respectively.
		\item[\textbf{Step 4}] Define the response function as the diffusion signal along directions in the y-z plane under a single tensor model with $\mathbf{D}=\rm{diag}\{\underline{\lambda}, \underline{\lambda}, \bar{\lambda}\}$:
		\[
			R(\cos(\theta)):= S_0\exp^{-b (\bar{\lambda} \cos^2\theta + \underline{\lambda} \sin^2\theta)}, ~~~ \theta \in  [0,\pi]
		\]

	\end{itemize}

	Note that, in our implementation, we first normalize the DWI measurements at each voxel by the mean intensity of the 6 $S_0$ images at that voxel. We then set $S_0=1$ in the response function. Since $S_0$ corresponds to a multiplicative factor in the response function SH coefficients matrix $\mathbf{R}$, such a normalization would not affect the fitted FOD.

	\subsection{Additional plots of the HCP D-MRI application}
	Here we provide additional plots of the HCP D-MRI application. 
	\begin{figure}[H]
		\centering
		\includegraphics[width=\textwidth]{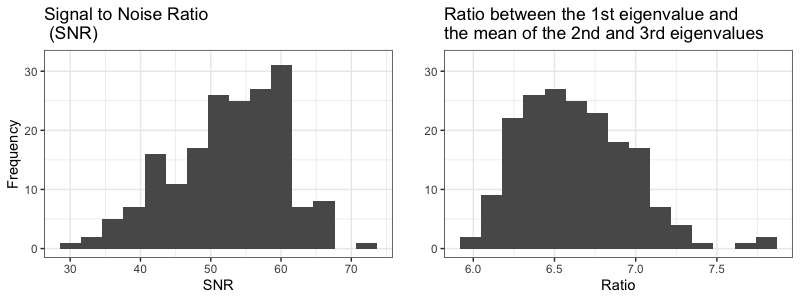}
		\caption{\textbf{Distributions of  (estimated) SNR and response function: } across $184$ sampled HCP subjects}
		\label{fig:snr}
	\end{figure}
	
	\begin{figure}[H]
		\centering
		\includegraphics[width=0.85\textwidth]{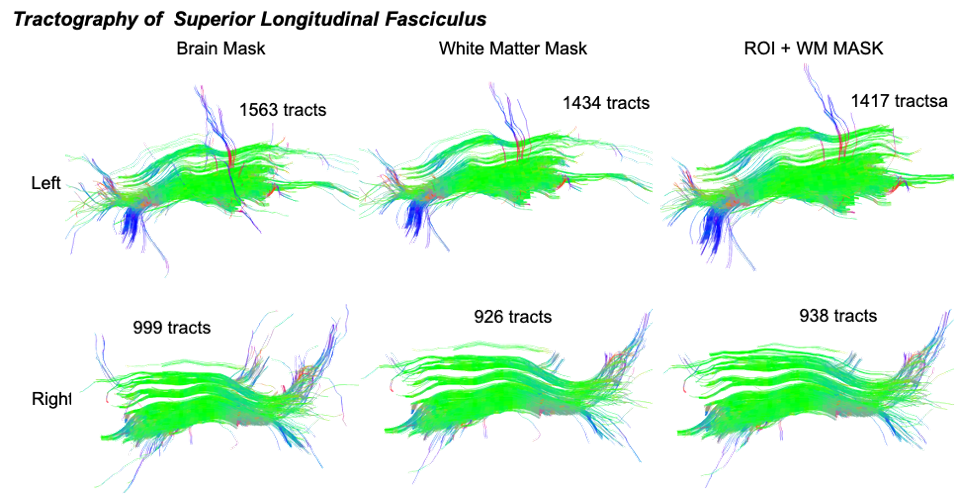}
		\caption{\textbf{SLF reconstruction of one HCP subject under different seeding strategies}. Left panel: whole-brain-seeding; Middle panel: white-matter seeding; Right panel: regional-seeding with white matter mask.  (color scheme: green: anterior-posterior; blue: superior-inferior; red: left-right)}
		\label{fig:slf_seeding}
	\end{figure}

	\begin{figure}[H]
		\centering
		\includegraphics[width=0.85\textwidth]{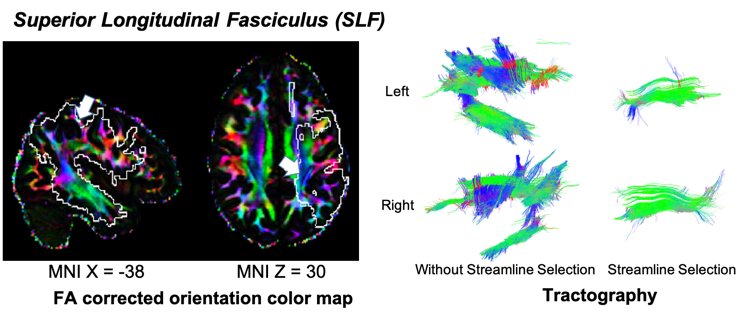}
		\caption{\textbf{FA corrected orientation color map of one HCP subject and the tractography of SLF}: Left Panel -- The (probabilistic) SLF mask on the left-hemisphere is outlined by white-colored lines with saggital view at MNI X = -38 and axial view at MNI Z = 30. Right Panel -- SLF tractography before and after streamline selection.  (color scheme: green: anterior-posterior; blue: superior-inferior; red: left-right)}
		\label{fig:colormap}
	\end{figure}

	\begin{figure}[H]
		\centering
		\includegraphics[width=0.5\textwidth]{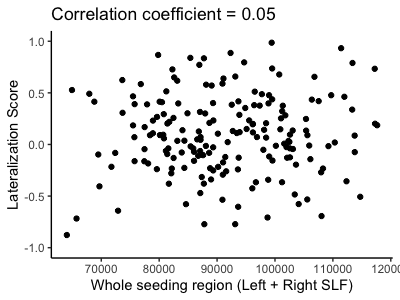}
		\caption{\textbf{Lateralization score vs. number of voxels in SLF ROIs:} across 184 sampled HCP subjects}
		\label{fig:scatter}
	\end{figure}

	\begin{figure}[H]
		\centering
		\includegraphics[width=0.9\textwidth]{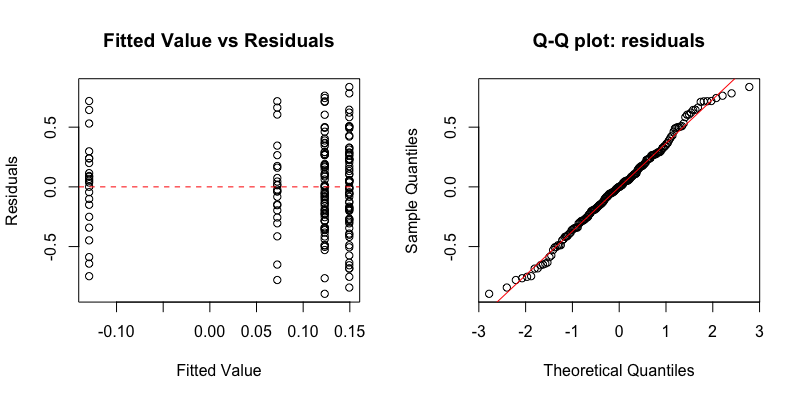}
		\caption{\textbf{HCP D-MRI application:} ANOVA model diagnostic plots}
		\label{fig:diagnostic}
	\end{figure}
	
	\subsection{HCP D-MRI application with DSI Studio}
	\label{sec:supp-dsi}	
	The steps we used in \textit{DSI Studio} are as follows:
	\begin{itemize}
		\item{\textbf{Step 1}} Convert the raw image to the input type of \textit{DSI Studio}.
		\item{\textbf{Step 2}} Use GQI (an ODF estimator) as the local fiber estimation method.
		\item{\textbf{Step 3}} Apply the deterministic tractography algorithm on SLF from the \textit{DSI Studio} Atlas.
		\item{\textbf{Step 4}} Calculate the lateralization score based on tractography results (i.e.,  number of fibers from left- and right- SLF). 
			\item{\textbf{Step 5}} Conduct two-way ANOVA to relate the lateralization score to gender and handedness.  
	\end{itemize}

\begin{figure}[H]
		\centering
		\includegraphics[width=\textwidth]{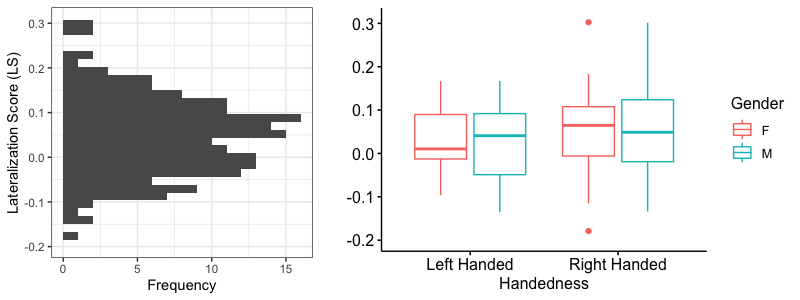}
		\caption{\textbf{Lateralization score distribution derived from \textit{DSI Studio}:} Left panel -- on  all subjects; Right panel -- by gender-handedness group}
		\label{fig:lateralization_score_dsi}
	\end{figure}

\textit{DSI Studio} gives  qualitatively similar results, even though the handedness effect is less significant (Fig.,  \ref{fig:lateralization_score_dsi}; Table  \ref{table:twowayanova_dsi}).

	\begin{table}[H]
	\caption{\textbf{HCP D-MRI application with \textit{DSI Studio}:} two-way ANOVA table}
	\label{table:twowayanova_dsi}
		\begin{tabular}{|c|c|c|c|c|c|}
		\hline
		\textbf{}                    & \textbf{d.f.} & \textbf{SS} & \textbf{MS} & \textbf{F-value} & \textbf{$p$-value} \\ \hline
		\textbf{Handedness}          & 1             & 0.0286       & 0.0286       & 3.585            & 0.0599             \\ \hline
		\textbf{Gender}              & 1             & 0.0001       & 0.0001       & 0.006            & 0.9367             \\ \hline
		\textbf{Handedness * Gender} & 1             & 0.0024       & 0.0024       & 0.299            & 0.5851             \\ \hline
		\textbf{Residuals}           & 180           & 1.4346      & 0.0080       &                  &                    \\ \hline
		\end{tabular}
	\end{table}

	\begin{figure}[H]
		\centering
		\includegraphics[width=0.9\textwidth]{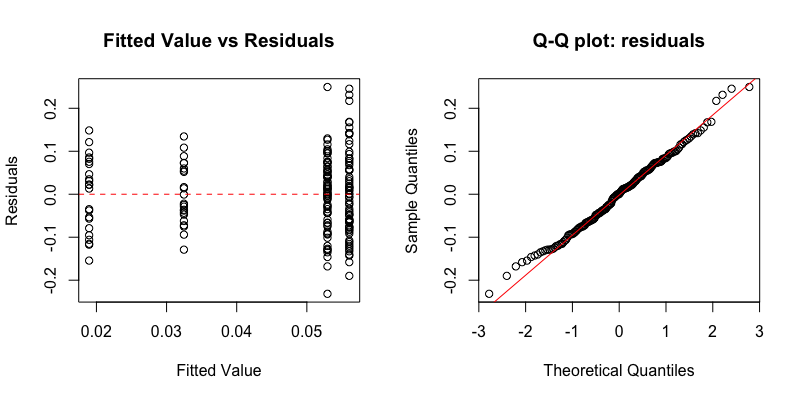}
		\caption{\textbf{HCP D-MRI application with \textit{DSI Studio}:} ANOVA model diagnostic plots}
		\label{fig:diagnostic_dsi}
	\end{figure}

\end{document}